\documentclass[useAMS,usenatbib]{mn2e}
\usepackage{epsfig}
\usepackage{amssymb}
\usepackage{amsmath}
\usepackage{lscape}
\usepackage{longtable}
\usepackage[normal]{caption}
\usepackage{multirow}

\title[\textit{Herschel}/HerMES: The X-ray -- Infrared correlation for star-forming galaxies at $z \sim$\,1]{\textit{Herschel}/HerMES: The X-ray -- Infrared Correlation for Star-forming Galaxies at z$\sim$\,1}

\author[M.~Symeonidis et al.]
{\parbox{\textwidth}{\raggedright M.~Symeonidis,$^{1}$\thanks{E-mail: \texttt{msy@mssl.ucl.ac.uk}}
A.~Georgakakis,$^{2}$
N.~Seymour,$^{1}$
R.~Auld,$^{3}$
J.~Bock,$^{4,5}$
D.~Brisbin,$^{6}$
V.~Buat,$^{7}$
D.~Burgarella,$^{7}$
P.~Chanial,$^{8}$
D.L.~Clements,$^{9}$
A.~Cooray,$^{10,4}$
S.~Eales,$^{3}$
D.~Farrah,$^{11}$
A.~Franceschini,$^{12}$
J.~Glenn,$^{13}$
M.~Griffin,$^{3}$
E.~Hatziminaoglou,$^{14}$
E.~Ibar,$^{15}$
R.J.~Ivison,$^{15,16}$
A.M.J.~Mortier,$^{9}$
S.J.~Oliver,$^{11}$
M.J.~Page,$^{1}$
A.~Papageorgiou,$^{3}$
C.P.~Pearson,$^{17,18}$
I.~P{\'e}rez-Fournon,$^{19,20}$
M.~Pohlen,$^{3}$
J.I.~Rawlings,$^{1}$
G.~Raymond,$^{3}$
G.~Rodighiero,$^{12}$
I.G.~Roseboom,$^{11}$
M.~Rowan-Robinson,$^{9}$
Douglas~Scott,$^{21}$
A.J.~Smith,$^{11}$
K.E.~Tugwell,$^{1}$
M.~Vaccari,$^{12}$
J.D.~Vieira,$^{4}$
L.~Vigroux,$^{22}$
L.~Wang$^{11}$ and
G.~Wright$^{15}$}\vspace{0.4cm}\\
\parbox{\textwidth}{\raggedright $^{1}$Mullard Space Science Laboratory, University College London, Holmbury St. Mary, Dorking, Surrey RH5 6NT, UK\\
$^{2}$National Observatory of Athens, Institute of Astronomy, V. Paulou \& I. Metaxa, Athens 15236, Greece\\
$^{3}$Cardiff School of Physics and Astronomy, Cardiff University, Queens Buildings, The Parade, Cardiff CF24 3AA, UK\\
$^{4}$California Institute of Technology, 1200 E. California Blvd., Pasadena, CA 91125, USA\\
$^{5}$Jet Propulsion Laboratory, 4800 Oak Grove Drive, Pasadena, CA 91109, USA\\
$^{6}$Space Science Building, Cornell University, Ithaca, NY, 14853-6801, USA\\
$^{7}$Laboratoire d'Astrophysique de Marseille, OAMP, Universit\'e Aix-marseille, CNRS, 38 rue Fr\'ed\'eric Joliot-Curie, 13388 Marseille cedex 13, France\\
$^{8}$Laboratoire AIM-Paris-Saclay, CEA/DSM/Irfu - CNRS - Universit\'e Paris Diderot, CE-Saclay, pt courrier 131, F-91191 Gif-sur-Yvette, France\\
$^{9}$Astrophysics Group, Imperial College London, Blackett Laboratory, Prince Consort Road, London SW7 2AZ, UK\\
$^{10}$Dept. of Physics \& Astronomy, University of California, Irvine, CA 92697, USA\\
$^{11}$Astronomy Centre, Dept. of Physics \& Astronomy, University of Sussex, Brighton BN1 9QH, UK\\
$^{12}$Dipartimento di Astronomia, Universit\`{a} di Padova, vicolo Osservatorio, 3, 35122 Padova, Italy\\
$^{13}$Dept. of Astrophysical and Planetary Sciences, CASA 389-UCB, University of Colorado, Boulder, CO 80309, USA\\
$^{14}$ESO, Karl-Schwarzschild-Str. 2, 85748 Garching bei M\"unchen, Germany\\
$^{15}$UK Astronomy Technology Centre, Royal Observatory, Blackford Hill, Edinburgh EH9 3HJ, UK\\
$^{16}$Institute for Astronomy, University of Edinburgh, Royal Observatory, Blackford Hill, Edinburgh EH9 3HJ, UK\\
$^{17}$Space Science \& Technology Department, Rutherford Appleton Laboratory, Chilton, Didcot, Oxfordshire OX11 0QX, UK\\
$^{18}$Institute for Space Imaging Science, University of Lethbridge, Lethbridge, Alberta, T1K 3M4, Canada\\
$^{19}$Instituto de Astrof{\'\i}sica de Canarias (IAC), E-38200 La Laguna, Tenerife, Spain\\
$^{20}$Departamento de Astrof{\'\i}sica, Universidad de La Laguna (ULL), E-38205 La Laguna, Tenerife, Spain\\
$^{21}$Department of Physics \& Astronomy, University of British Columbia, 6224 Agricultural Road, Vancouver, BC V6T~1Z1, Canada\\
$^{22}$Institut d'Astrophysique de Paris, UMR 7095, CNRS, UPMC Univ. Paris 06, 98bis boulevard Arago, F-75014 Paris, France}}

\begin{document}

\date{Accepted  Received; in original form}

\pagerange{\pageref{firstpage}--\pageref{lastpage}} \pubyear{2010}

\maketitle

\label{firstpage}

\begin{abstract}
For the first time, we investigate the X-ray/infrared (IR) correlation for star-forming galaxies at $z \sim$\,1, using SPIRE submm data from
the recently-launched \textit{Herschel Space Observatory} and deep X-ray data from the 2\,Ms \textit{Chandra} deep field north (CDFN)
survey. We examine the X-ray/IR correlation in the soft X-ray (SX, 0.5--2\,keV) and hard X-ray (HX, 2--10\,keV) bands by comparing our $z
\sim$\,1 SPIRE-detected star-forming galaxies (SFGs) to equivalently IR-luminous ($L_{\rm IR}$\,$>$\,10$^{10}$\,L$_{\odot}$) samples in the
local/low redshift Universe. Our results suggest that the X-ray/IR properties of the SPIRE SFGs are on average similar to those of their
local counterparts, as we find no evidence for evolution in the $L_{\rm SX}$/$L_{\rm IR}$ and  $L_{\rm HX}$/$L_{\rm IR}$ ratios with
redshift. We note however, that at all redshifts, both $L_{\rm SX}$/$L_{\rm IR}$ and  $L_{\rm HX}$/$L_{\rm IR}$ are strongly
dependent on IR luminosity, with luminous and ultraluminous infrared
galaxies (LIRGs and ULIRGs, $L_{\rm IR}$\,$>$\,10$^{11}$\,L$_{\odot}$)
having up to an order of magnitude lower values than normal infrared
galaxies ($L_{\rm IR}$\,$<$\,10$^{11}$\,L$_{\odot}$). 
We derive a $L_{\rm SX}$-$L_{\rm IR}$ relation and confirm the applicability of an existing $L_{\rm HX}$-$L_{\rm IR}$
relation for both local and distant LIRGs and ULIRGs, consistent with a scenario where X-ray luminosity is correlated with the star-formation
rate (SFR). 

\end{abstract}


\section{Introduction}

For at least two decades, the infrared/X-ray connection has been
an important point of focus in extragalactic surveys. The various X-ray missions (\textit{Einstein}, \textit{ROSAT}, \textit{ASCA},
\textit{BeppoSAX}, \textit{Chandra} and \textit{XMM-Newton}) together with the infrared observatories (\textit{IRAS}, \textit{ISO} and
\textit{Spitzer}) have established the X-ray/infrared synergy not only as a tool for studying the active galactic nucleus/host galaxy
interplay, but also star-formation and stellar evolution (a few examples of the multitude of papers on this topic: Eales $\&$ Arnaud
1988\nocite{EA88}; Fabbiano 1988\nocite{Fabbiano88}; Fabbiano, Gioia $\&$ Trinchieri 1988\nocite{FGT88}; Rieke 1988\nocite{Rieke88}; Green
et al. 1989\nocite{Green89}; Rephaeli et al. 1991\nocite{Rephaeli91}; Green et al. 1992\nocite{Green92}; Boller et
al. 1992\nocite{Boller92}; Barcons et al. 1995\nocite{Barcons95}; Rephaeli, Gruber $\&$ Persic 1995\nocite{RGP95}; Imanishi $\&$ Ueno
1999\nocite{IU99}; Severgnini et al. 2000\nocite{Severgnini00}; Risaliti et al. 2000\nocite{Risaliti00}; Cavaliere, Giacconi $\&$
Menci 2000\nocite{CGM00}; Fadda et al. 2002\nocite{Fadda02}; Gandhi $\&$ Fabian 2003\nocite{GF03}; Manners et al. 2004\nocite{Manners04};
Panessa et al. 2005\nocite{Panessa05}; Netzer et al. 2005\nocite{Netzer05}; Franceschini et
al. 2005\nocite{Franceschini05}; Hickox et al. 2007\nocite{Hickox07}; Treister et al. 2009\nocite{Treister09}; Trichas et al. 2009\nocite{Trichas09}; Lutz et al. 2010\nocite{Lutz10}; Park et
al. 2010\nocite{Park10}; Goulding et al. 2011\nocite{Goulding11}).

Emission in the infrared (IR) to millimetre ($\sim$\,8--1000\,$\mu$m) part of the spectrum is attributed to galactic dust being heated to a range
of temperatures by incident UV and optical photons. In cases where an active galactic nucleus (AGN) is present, emission from the torus (or other axisymmetric
dust/gas configurations, e.g. Efstathiou $\&$ Rowan-Robinson 1995\nocite{ERR95}) can also play a prominent role at short infrared
wavelengths ($<$\,20\,$\mu$m, rest-frame), as torus dust is heated to near-sublimation temperatures by the AGN radiation field. On the other
hand, flux longwards of $\sim$\,20\,$\mu$m is strongly correlated with the star formation rate (SFR), although in the case of quiescently
star-forming galaxies, emission from dust heated by older stellar populations might also contribute significantly to the far-IR (e.g. Bendo
et al. 2010\nocite{Bendo10}, Calzetti et al. 2010\nocite{Calzetti10}). Nevertheless, in galaxies with high total infrared luminosities ($L_{\rm IR}$\,$>$10$^{10}$\,L$_{\odot}$)
where large dust masses are implied, (i) the source of heating is intense and (ii) the dust replenishment is continuous, making IR
emission an excellent tracer of young star-forming/starburst regions. Moreover, contamination from an AGN (if present) is minimal
in the far-IR/submm part of the spectrum. Recently, Hatziminaoglou et al. (2010\nocite{Hatziminaoglou10}) showed that the submm colours of
AGN hosts are indistinguishable from those of the general submm-detected population. 

Emission in the high energy part of the spectrum is a combination of both AGN (if present) and star-formation related processes. AGN are
the most powerful X-ray emitters, with X-rays originating in the hot
corona surrounding the accretion disk through inverse Compton scattering of
softer energy photons (Sunyaev $\&$ Titarchuk 1980\nocite{ST80}),
generally amounting to a few per cent of the AGN energy budget (Elvis
1994\nocite{Elvis94}). On the other hand, the applicability of X-rays as a star formation tracer is linked to a galaxy's star formation
history and interstellar medium (ISM) conditions (e.g. Ghosh $\&$ White 2001\nocite{GW01}; Fabbiano et
al. 2004\nocite{Fabbiano04}). Star-forming galaxies have complex 0.5--10\,keV spectra, with contributions from a low temperature soft
component (k$T\lesssim$1\,keV) and a harder component which is either predominantly thermal (k$T\sim$5--10\,keV) or non-thermal (power-law
index $\Gamma$\,$\sim$1.5--2) or a combination of both (Persic $\&$
Rephaeli 2002\nocite{PR02}). The soft component is primarily thermal
emission from gas in the ISM heated to X-ray temperatures by stellar
winds and supernovae, with a contribution from point sources of
the order of 30 per cent, according to Chandra observations of local
galaxies (e.g. Fabbiano, Zezas $\&$ Murray 2001\nocite{FZM01}; Zezas,
Ward $\&$ Murray 2003\nocite{ZWM03}). On the other hand,
more than 60 per cent of the hard component is
resolved into point sources and associated with X-ray binaries (e.g.
Griffiths et al. 2000\nocite{Griffiths00}).  High mass X-ray binary (HMXB) spectra can be represented by a power law up to about 20\,keV, whereas the
spectra of low mass X-ray binaries (LMXB) are a combination of thermal bremmstrahlung and black-body emission in the $<$20\,keV region
(e.g. Christian $\&$ Swank 1997\nocite{CS97}, see also Persic $\&$ Rephaeli 2002).
HMXBs, where the companion star is massive and short-lived, are direct tracers of recent star-formation, whereas LMXBs have lifetimes of the
order of 10\,Gyr and are hence more appropriate tracers of stellar mass (e.g. Ptak et al. 2001\nocite{Ptak01}; Grimm, Gilfanov $\&$
Sunyaev 2002, 2003\nocite{GGS02}\nocite{GGS03}). Although the relative timescales of X-ray emitting processes play an important role in how
one interprets the star-formation history of quiescently star-forming galaxies or starburst (SB) systems, X-rays are a good overall tracer
of the first $\sim$\,30\,Myr of star-forming activity (e.g. Mass-Hesse, Oti-Floranes $\&$ Cervino 2008\nocite{MOC08}), with
the relation between star formation rate (SFR) and full band X-ray luminosity found to be linear in the $>$1\,M$_{\odot}\,\rm yr^{-1}$
SFR regime (Gilfanov, Grimm $\&$ Sunyaev 2004\nocite{GGS04}).

As both X-ray and IR emission trace star formation, one would expect these quantities to be correlated in the absence of an AGN. The
X-ray/IR correlation, described as log\,($L_{\rm  X}$)=A\,+\,B\,log\,($L_{\rm IR}$), has been studied extensively in both the soft (0.5--2\,keV) and hard (2--10\,keV) energy bands with
samples of local/low-redshift galaxies (eg. Griffiths $\&$ Padovani 1990\nocite{GP90}; David, Jones $\&$ Forman 1992\nocite{DJF92};
Ranalli, Comastri $\&$ Setti 2003\nocite{RCS03}; Franceschini et al. 2003\nocite{Franceschini03b}; Rosa-Gonz{\'a}lez et
al. 2007). These studies have calculated different values for the slope of the soft X-ray/IR relation with B roughly ranging from 0.7 to 1. A
linear relation (B=1) is a strong indication that X-ray and IR emission directly trace the star formation rate, however deviations from
linearity do not necessarily suggest the contrary. The large range in computed slopes could be a consequence of the small number statistics in the
samples studied, in combination with the large scatter in the galaxies' physical properties, such as the efficiency of energy coupling to the
ISM which heats the X-ray emitting gas, varying contributions from
point sources, as well as the amount of absorption undergone by the soft X-rays due to gas/dust columns in the
line of sight. In that respect, hard X-rays are a more reliable tracer of star-formation, as the ISM is mostly transparent at energies above
\,$\sim$\,2keV, except for dense, massive molecular cloud environments (Grimm, Gilfanov $\&$ Sunyaev 2002; Ranalli, Comastri $\&$ Setti 2003;
Persic et al. 2004\nocite{Persic04}). Indeed, hard X-ray emission linearly traces the star formation rate (e.g. Grimm, Gilfanov $\&$
Sunyaev 2003; Persic et al. 2004), although because of the lower sensitivity of detectors in the hard X-rays, this band is still not as
extensively used. In addition, whereas soft X-rays suffer little contamination from X-ray binary emission, to use hard X-rays successfully, the relative contribution of HMXBs and LMXBs needs to be
quantified. This is easier for SB environments where HMXBs dominate but for more quiescently star-forming systems, LMXBs, which trace the
stellar mass, can contribute substantially, introducing non-linearities (e.g. Grimm, Gilfanov $\&$ Sunyaev 2002; Persic et al
2004\nocite{Persic04}). 

So, what is the role of the X-ray/IR correlation in extragalactic surveys? Apart from the obvious motivation, i.e. examining the
physical properties of different galaxy populations, one of its key applications could be to discriminate between AGN-dominated and
star-formation dominated systems. The ability to discriminate between the two is particularly significant in X-ray surveys,
where AGN dominate the bright end of the number counts, but the star-forming galaxy population is expected to outnumber AGN at faint
fluxes (f$_{\rm 0.5–-2\,keV}$\,$\sim$\,10$^{-17}$\,erg\,s$^{-1}$\,cm$^{-2}$; e.g. Hornschemeier et al. 2002\nocite{Hornschemeier02}, 2003\nocite{Hornschemeier03}; Bauer et al. 2004; Brandt $\&$ Hasinger
2005\nocite{BH05}; see also Georgakakis et al. 2003\nocite{Georgakakis03}; 2004\nocite{Georgakakis04}; 2006ab\nocite{Georgakakis06a}\nocite{Georgakakis06b};
2007\nocite{Georgakakis07}; 2008\nocite{Georgakakis08}). As previously shown (e.g. Georgakakis et al. 2007\nocite{Georgakakis07}), the
X-ray/IR correlation is a useful way of separating starbursts from AGN. However, up to now, such studies have used relations applicable
to normal IR galaxies in the local Universe. What remains to be settled is whether there is a change in the characteristics of the
X-ray/IR relation with redshift and/or SFR. As noted earlier, a linear relationship for star-forming galaxies is a strong indication that
X-ray and IR emission directly trace the star formation rate. As a result, determining the X-ray/IR relation would play an important role
in calibrating X-ray emission as a star formation indicator, providing an alternative census of the global SFR density. 

Although various extragalactic X-ray studies have targetted cosmologically significant redshifts (e.g. Norman et al. 2004\nocite{Norman04}; Ptak
et al. 2007\nocite{Ptak07}; Georgakakis et al 2007; Lehmer et al. 2008\nocite{Lehmer08}), the X-ray/IR correlation has been examined only for local galaxies (e.g. Ranalli, Comastri $\&$
Setti et al. 2003) and X-ray selected samples at moderate redshifts (Grimm, Gilvanov $\&$ Sunyaev 2002; Rosa-Gonz{\'a}lez et
al. 2007\nocite{Rosa-Gonzalez07}), mainly due to sensitivity limitations in X-ray surveys.
For the first time, we are able to investigate the X-ray/IR correlation at $z$\,$\sim$1, for submm-bright sources with high
star-formation rates, using deep X-ray data from the \textit{Chandra} 2\,Ms CDFN survey and sub-mm data from SPIRE (Griffin et al. 2010\nocite{Griffin10}), the sub-mm bolometer array on the recently-launched
\textit{Herschel Space Observatory}\footnote{\textit{Herschel} is an ESA space observatory with science instruments provided by European-led
  Principal Investigator consortia and with important participation from NASA.} (Pilbratt et al. 2010\nocite{Pilbratt10}). Our aim is to compare the X-ray properties of $z$\,$\sim$1 IR-luminous
galaxies to their counterparts in the local ($z<0.1$), moderate
($z\sim0.6$) and high ($z\sim2$) redshift Universe. The paper is laid
out as follows: in Section \ref{sec:sample} we present the sample and in Sections \ref{sec:soft} and \ref{sec:hard}
we examine the soft and hard X-ray/IR relations. Our summary and conclusions are presented in Section \ref{sec:conclusions}. Throughout this paper, we adopt a concordance consmology of H$_0$=70\,km\,s$^{-1}$Mpc$^{-1}$, $\Omega_{\rm M}$=1-$\Omega_{\rm \Lambda}$=0.3.

\section{The data and sample selection}
\label{sec:sample}

\subsection{Optical and Infrared data}
\label{sec:IRdata}

The infrared data come from the \textit{Herschel}/SPIRE Science Demonstration Phase (SDP) observations of the Great Observatories Origins Deep Survey (GOODS)-North, taken as part of the \textit{Herschel} Multi-tiered Extragalactic Survey (HerMES) guaranteed-time key programme\footnote{hermes.sussex.ac.uk} (Oliver et al. in prep.). We focus on the region of overlap (0.08 deg$^2$) between the HerMES GOODS-N coverage and the deep 24\,$\mu$m \textit{Spitzer}/MIPS (Rieke et al. 2004\nocite{Rieke04}) observations (Magnelli et al. 2009). 
Source extraction on \textit{Herschel}/SPIRE 250, 350 and 500\,$\mu$m images\footnote{The data presented in this paper will be released through the {\em Herschel} Database in Marseille HeDaM
  ({hedam.oamp.fr/HerMES})} is performed on the IRAC-3.6\,$\mu$m positions of the 5\,$\sigma$ 24\,$\mu$m GOODS-N sources as described in Roseboom et al. (2010\nocite{Roseboom10}).
We find 84 sources with $f_{250\mu m}$ above the 3\,$\sigma$ confusion of 17.4 mJy (see Nguyen et al. 2010\nocite{Nguyen10} for the SPIRE confusion limits), hereafter referred to as the full SPIRE sample. This method of source extraction on already known positions is widely used and enables identifications of secure counterparts over the whole spectral energy distribution (SED). In this case however, its significant advantage, lies in its ability to effectively deal with source blending in the SPIRE bands. Due to the large SPIRE beams of 18.1, 24.9 and 36.6 arcsec (FWHM) at 250, 350 and 500\,$\mu$m respectively (Nguyen et al. 2010), many SPIRE detections are composed of confused sources. By using IRAC-3.6\,$\mu$m positions of 24$\mu$m sources, we are able to assign SPIRE fluxes to each 24$\mu$m source in the blended composit and hence obtain the aforementioned `clean' full SPIRE sample of 84 sources.
A detailed report on the reliability and completeness of this method is presented in Roseboom et al. (2010), where it is judged in the context of results from the sub-mm number counts and cosmic infrared background. Roseboom et al. demonstrate that for deep 24\,$\mu$m surveys and bright SPIRE flux densities, the 250\,$\mu$m-detected, 24\,$\mu$m sources are essentially the objects that make up the 250\,$\mu$m population. 

Following the same method of source extraction as for the SPIRE bands we also obtain photometry at \textit{Spitzer}/MIPS 70 and 160\,$\mu$m in order to better sample the infrared SED. We also retrieve R-band magnitudes, by crossmatching the IRAC-3.6\,$\mu$m positions of the full SPIRE sample with the Capak et al. (2004)\nocite{Capak04} GOODS-N optical catalogue, within a radius of 1.5 arcsec. The association rate is 90 per cent.

\begin{table*}
\centering
\caption{Table of stacking results for the X-ray non-detections in the redshift sub-sample of 67 sources, grouped into LIRGs and ULIRGs. Columns 2 and 3 show the mean total infrared luminosity (with 1\,$\sigma$ uncertainties) and mean redshift of the group. Columns 4--7 display information for the soft and hard band respectively. $f_{\rm SX}$ and $f_{\rm HX}$ (erg\,s$^{-1}$\,cm$^{-2}$) are the soft and hard stacked fluxes of each group. The significance of each stacked signal is shown in brackets, where $\sigma$ refers to the background level. Corresponding luminosities $L_{\rm SX}$ and $L_{\rm HX}$ (erg\,s$^{-1}$) are estimated using the mean redshift of the group. Note that for the ULIRGs, there is no stacked signal and therefore we quote the 3$\sigma$ upper limit. }
\begin{tabular}{l|l|l|l|l|l|l|}
\hline 
& &&\multicolumn{2}{|c|}{Soft band: 0.5--2\,keV} & \multicolumn{2}{|c|}{Hard band: 2--10\,keV} \\ 
Group& log\,$\left\langle L_{\rm IR} \right\rangle$ &$\left\langle z \right\rangle$ & $f_{\rm SX}$  & log\,$L_{\rm SX}$  &$f_{\rm HX}$  & log\,$L_{\rm HX}$  \\
\hline
LIRGs (18 sources) &11.55$^{+0.23}_{-0.51}$&0.62&2.37($\pm$0.45)$\times$10$^{-17}$ &40.56$^{+0.08}_{-0.09}$&7.08($\pm$2.5)$\times$10$^{-17}$ &41.04$^{+0.13}_{-0.2}$\\
& &&(9.3\,$\sigma$) &&(4.3\,$\sigma$) &\\
ULIRGs (5 sources) &12.34$^{+0.18}_{-0.32}$&1.34&2.43($\pm$0.74) $\times$10$^{-17}$ &41.38$^{+0.11}_{-0.16}$&$<$8.03 $\times$10$^{-17}$&$<$41.9\\
& &&(6\,$\sigma$) &&($<$3\,$\sigma$) &\\

\hline
\end{tabular}
\label{table:stack}
\end{table*}

\subsection{X-ray data}
\label{sec:xraydata}

The X-ray observations are from the 2\,Ms \textit{Chandra} Deep Field North (CDFN) survey.  The methodology for data reduction, source detection and photometry estimates is described in Laird et
al. (2009\nocite{Laird09}) and the final data products are made available by the Imperial College (IC) team\footnote{http://astro.ic.ac.uk/research/data-products-chandra-surveys}. The IC catalogue consists of X-ray sources with a Poisson significance of $<4\times10^{-6}$ (equivalent to $>$\,4.5$\sigma$ in the case of a normal distribution) detected independently in four energy bands, full (0.5--7\,keV), soft (0.5--2\,keV), hard (2--7\,keV) and ultra-hard (5-7\,keV). Soft and hard X-ray fluxes are estimated in the 0.5--2 and 2--10\,keV energy intervals, adopting a photon index of $\rm \Gamma=1.4$ and using the Bayesian methodology described in Laird et al. (2009). We cross-match the IC catalogue to the IRAC-3.6\,$\mu$m coordinates of the full SPIRE sample, within a 2\,arcsec radius and find that 37 out of 84 sources can be matched with a formal X-ray counterpart in at least one of the soft or hard X-ray bands (44 per cent). 

In order to reach higher completeness with respect to the X-ray association rate, we opt to go deeper by using the IRAC-3.6\,$\mu$m position of each SPIRE source to extract counts from the X-ray map down to a Poisson significance $<3\times10^{-3}$ (corresponds to $>$\,3$\sigma$ in the case of a normal distribution). For each object, the X-ray counts
within the 70 per cent encircled energy fraction (EEF) radius are summed up and a local value for
the background is estimated using an annulus centered on the source with an inner radius 1.5 times the 90 per cent EEF and a width of
50\,arcsec (100 pixels).  For the estimation of the background, X-ray sources with Poisson significance $<4\times10^{-6}$ are removed by
excluding pixels within the 95 per cent EEF radius of each source. The probability that the observed counts are a random fluctuation of the
background is estimated using Poisson statistics. As in the IC catalogue, fluxes are estimated using a Bayesian methodology similar to that described in Laird et al. (2009) by integrating the source counts within the 90 per cent EEF aperture. The count rates of sources in the soft and hard bands are converted to fluxes (erg\,s$^{-1}$\,cm$^{-2}$) in the 0.5--2 and
2--10\,keV energy intervals respectively by adopting a power-law X-ray spectrum with index $\rm \Gamma=1.9$, absorbed by the Galactic gas
column density in the direction of the CDFN field, $N_{\rm H}$=2$\times$10$^{20}$\,cm$^{-2}$. The uncertainties on the fluxes correspond to the 68 per cent confidence level.  We choose $\rm \Gamma=1.9$ rather than the $\rm \Gamma=1.4$ used in the IC catalogue, as the former is
consistent with the typically soft X-ray spectra of star-forming galaxies and unabsorbed AGN (e.g. Nandra $\&$ Pounds
1994\nocite{NP94}) and hence more appropriate for our study. 
The X-ray detection rate now increases to 55 per cent (46 out of 84 objects have an X-ray flux of $>$\,3$\sigma$ in at least one X-ray band).
For objects without a formal or low significance X-ray counterpart, the 99.7 per cent confidence upper limit for their fluxes is
estimated. 

We also estimate hardness ratios (HRs), defined as $(H-S)/(H+S)$ where $H$ and $S$ refer to the hard (2--7\,keV) and soft (0.5--2\,keV) count rates (counts\,s$^{-1}$). The hardness ratio describes the relative strength of the hard to soft band signal and hence gives an indication of the shape of the X-ray spectrum. 
The X-ray properties of the sample are displayed in table \ref{table:all}.

\subsection{The SPIRE sample}
\label{sec:full_sample}

\begin{figure}
\epsfig{file=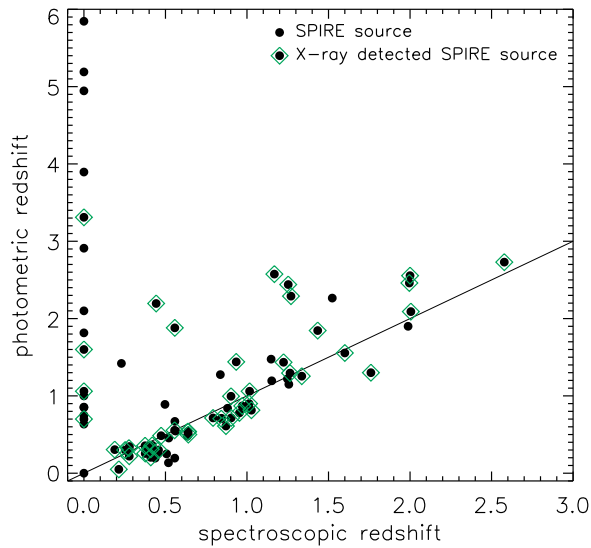,width=9cm}
\caption{Photometric versus spectroscopic redshifts for the full SPIRE sample of 84 sources. The solid
  line represents a one-to-one relationship. The 43 X-ray detected objects are denoted with green diamonds. All objects at (x,y) = (0,p) have no
  spectroscopic redshift and a photometric redshift of p. Note that
  most photometric redshifts above z$\sim$1, deviate significantly from the true redshift.}
\label{fig:redshifts}
\end{figure}

In the full SPIRE sample, 67 out of 84 sources (80 per cent) are associated with a spectroscopic redshift and 16 with a photometric redshift only (1 has neither a spectroscopic nor a photometric redshift) --- see Raymond et al. in prep.; also Eales et al. 2010\nocite{Eales10} for photometric redshifts. The spectroscopic redshifts come from the Barger, Cowie $\&$ Wang (2008\nocite{BCW08}) spectroscopic survey of the GOODS-N 24\,$\mu$m population, the survey being more than 90 per cent complete at f$_{24\,\mu m}$\,$>$250\,$\mu$Jy.
We investigate the reliability of the photometric redshifts and find that although there is good agreement below z$\sim$1, most of the $z>$1 photometric redshifts significantly deviate ($\Delta z>$0.5) from the true redshift (see Fig. \ref{fig:redshifts}). Hence, by extrapolation, the redshift estimates for the majority of our photometric sub-sample (11 objects with $z_{\rm phot}>$1) are unlikely to be reliable. 
We therefore choose to restrict the work presented in this paper to the 67 spectroscopically identified SPIRE sources. This excludes 4 X-ray detected sources, only 2 of which are potentially relevant to our study, the others having hard hardness ratios characteristic of AGN (see section \ref{sec:agn_fraction} on the selection of AGN). As a result, the 67 sources with spectroscopic redshifts constitute a near-complete sample for our purposes and hereafter, for simplicity, we refer to them as as the `SPIRE sample'.

The SPIRE sample of 67 sources spans a redshift range of 0.2\,$<z<$\,2.6, with a mean of $\left\langle z \right\rangle$\,=0.86. We calculate total infrared luminosities ($L_{\rm IR}$/L$_{\odot}$, 8--1000\,$\mu$m), by performing a $\chi^2$ fit on the available photometry (24--500\,$\mu$m), using the entire range of the Siebenmorgen $\&$ Kr{\"u}gel (2007, SK07)\nocite{SK07} library which consists of about 7000 templates. In Symeonidis et al. (2008)\nocite{Symeonidis08} we showed that the SK07 library can accommodate a large range of SEDs and multiple dust temperature components and is thus ideal for the dusty starburst sample we have selected here. Following the method outlined in Symeonidis et al. (2009\nocite{Symeonidis09}), we calculate the total infrared luminosity ($L_{\rm IR}$, 8--1000\,$\mu$m) and corresponding uncertainty for each object. The resulting 1$\sigma$ uncertainty in $L_{\rm IR}$ (typically of the order of 5 per cent) is combined with a systematic uncertainty of 8 per cent which takes into account the average calibration uncertainty of MIPS and SPIRE. The final fractional uncertainties on $L_{\rm IR}$ are comparable to the fractional uncertainties on the SPIRE fluxes. 
We find 38 sources to be in the luminous infrared galaxy regime (LIRGs, 57 per cent), 22 are ultraluminous infrared galaxies (ULIRGs, 33 per cent) and 7 are normal IR galaxies ($L_{\rm IR}$/L$_{\odot}$\,$<$\,10$^{11}$). 

To calculate X-ray luminosities, the soft and hard fluxes are \textit{K}-corrected using a photon index $\Gamma$=1.9; hereafter $L_{\rm SX}$ and $L_{\rm HX}$ refer to the soft (0.5--2\,keV) and hard (2--10\,keV) X-ray luminosities. For objects with X-ray upper limits, we use stacking analysis (described in Georgakakis et al. 2008) to explore their mean X-ray flux, by grouping them into LIRGs and ULIRGs and stacking separately in the soft and hard X-ray bands; see table \ref{table:stack} for the stacking results. For the normal IR galaxies we are unable to perform stacking as only 2 sources are not X-ray detected. X-ray luminosities for the stacked groups are computed using the mean redshift of the group. 

\begin{figure}
\epsfig{file=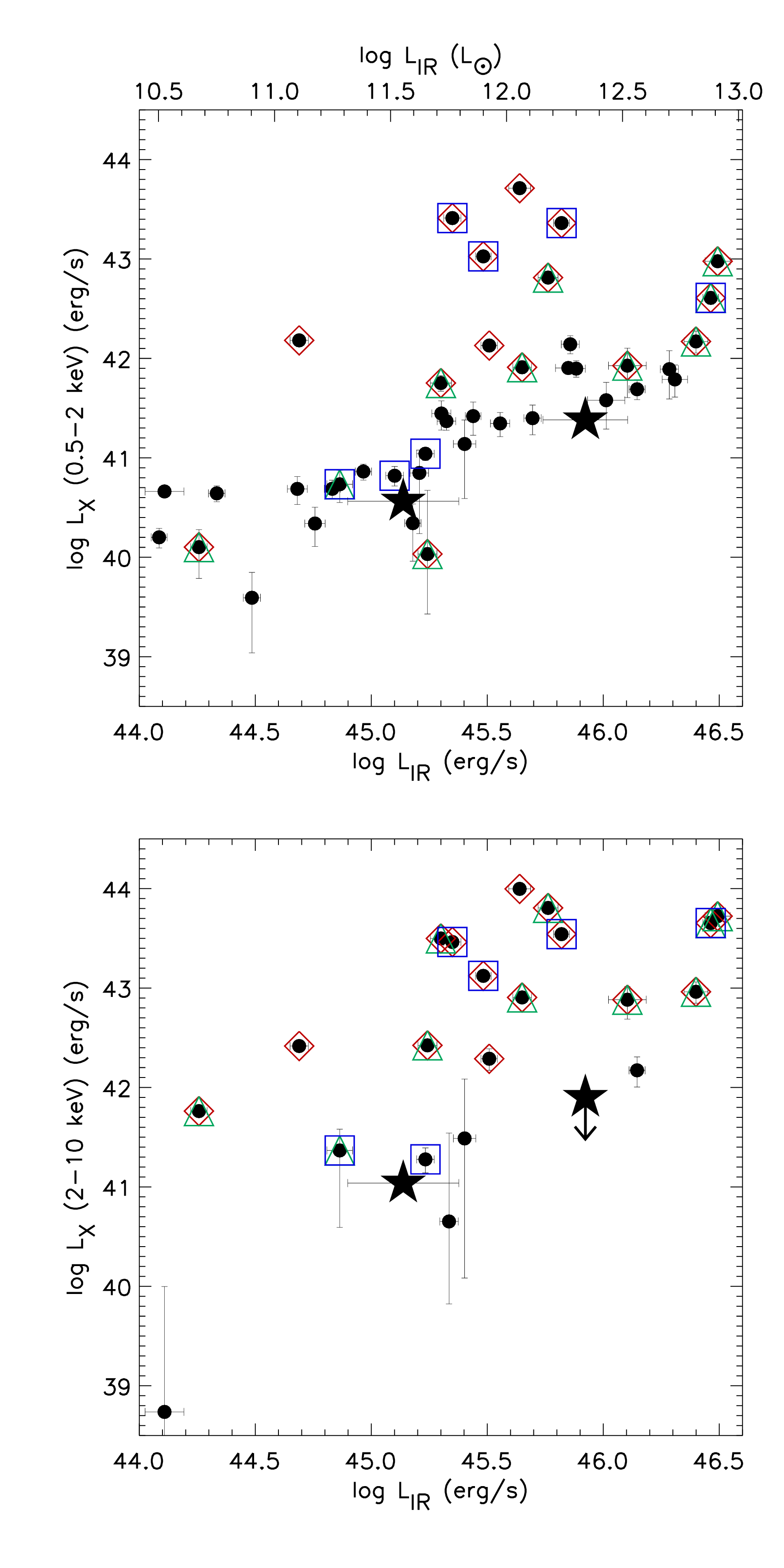,width=9.cm}
\caption{The X-ray versus IR luminosities of the 67 SPIRE sources with redshifts --- on the top panel we plot $L_{\rm SX}$ vs $L_{\rm IR}$ and on the lower panel $L_{\rm HX}$ vs $L_{\rm IR}$. X-ray detected sources are denoted by black filled circles. Non-detections are incorporated into the stacked groups denoted by the 2 large stars (see table \ref{table:stack} for details on the stacking). Note that for the ULIRG category there is no stacked signal in the hard band and we plot the 3\,$\sigma$ upper limit. AGN classified objects are marked on the plot by green triangles, red diamonds and blue squares, depending on the criterion used to identify them. Green triangles are hard sources with HR$>$-0.1, blue squares correspond to AGN identified through high excitation or broad lines in their optical spectra and red diamonds represent log\,[f$_{\rm HX}$/f$_{\rm R}$]\,$>$\,-1 sources. Note that all, bar one, log\,[f$_{\rm HX}$/f$_{\rm R}$]\,$>$\,-1 sources also have a hard X-ray luminosity greater than 10$^{42}$\,erg\,s$^{-1}$. The two LIRGs and one ULIRG which are hard X-ray detected, but have no associated AGN symbol, are ambiguous classifications (see Section \ref{sec:agn_fraction}). Although they are included in the star-forming sub-sample, our conclusions do not change if we include them in the AGN sub-sample instead. }
\label{fig:xray_ir}
\end{figure}

Fig. \ref{fig:xray_ir} shows the X-ray/IR luminosities of the SPIRE sample, with the AGN-classified sources clearly marked on the plot (see section \ref{sec:agn_fraction} for details on AGN classification). Note that due to the lower sensitivity of the hard X-ray band, many sources have no hard X-ray counterpart --- the soft X-ray association rate is 54 per cent whereas the hard X-ray association rate is 29 per cent.

\subsection{The AGN fraction}
\label{sec:agn_fraction}

As our analysis focuses on the X-ray/IR properties of star-forming galaxies, we require the candidate sources to be dominated by star-formation in both the X-rays and the infrared. The latter is ensured as our sample is submm-bright; as mentioned earlier, the dust masses implied for galaxies whose SEDs peak in the far-IR/submm, as well as their high energy output in that part
of the spectrum, suggests that infrared emission is strongly linked to star-formation. Weeding out AGN from the sample therefore rests on diagnosing the origin of the X-ray emission. 

We investigate the incidence of AGN with the aid of 2 X-ray criteria: the hard X-ray to R-band flux ratio (f$_{\rm HX}$/f$_{\rm R}$) and the hardness ratio, relying on the well established result that AGN are the most powerful X-ray emitters and that hard X-rays are minimally absorbed by the torus gas and dust. The f$_{\rm X}$/f$_{\rm R}$ ratio has been extensively used to separate
AGN and starburst systems since early observations of spectroscopically identified AGN have shown them to dominate the
-1$<$\,log\,[f$_{\rm X}$/f$_{\rm R}$]$<$1 parameter space (e.g. Schmidt et al. 1998\nocite{Schmidt98}; Akiyama et
al. 2000\nocite{Akiyama00}), with star-forming galaxies having values of log\,[f$_{\rm X}$/f$_{\rm R}$]$<$-1 and typically $<$-2 (e.g. Hornschemeier et al. 2001\nocite{Hornschemeier01}, 2003\nocite{Hornschemeier03}; Akiyama et al. 2003\nocite{Akiyama03}; Georgantopoulos, Georgakakis $\&$ Koulouridis 2005\nocite{GGK05};
Georgakakis et al. 2007\nocite{Georgakakis07}). Our second criterion (the hardness ratio) is related to the differences between observed X-ray spectra of AGN and star-forming galaxies. Although both star-forming systems and unabsorbed AGN are characterised by soft X-ray spectra and hence low hardness ratios, absorbed AGN have hard spectra and high values of hardness ratio, as low X-ray frequencies are more severely attenuated by gas column densities in the line of sight. Measured photon indices ($\Gamma$) of starburst galaxies range between 3 and 1.2 (e.g. Franceschini et al. 2003; Lehmer et al. 2010), corresponding to hardness ratios between -0.65 and -0.1 and therefore a cut at HR=-0.1 should successfully isolate absorbed AGN. 

Although the properties of star-forming galaxies have not been seen to deviate from log\,[f$_{\rm X}$/f$_{\rm R}$]$\lesssim$-1, low luminosity AGN do exhibit properties which overlap with the starburst parameter space and hence will be missed. In addition, as this indicator is linked to a hard band detection, AGN which are only soft-band detected will not be identified. Finally, the hardness ratio diagnostic neglects all unabsorbed AGN. Aiming for a higher completeness in the AGN fraction, we supplement our diagnosis with optical spectroscopy, through inspection of the spectral classification of the X-ray-detected SPIRE sources in Trouille et al. (2008\nocite{Trouille08}). In total, we find 25 X-ray sources with an assigned spectral class as defined in Barger et al. (2005); 5 are high-excitation-line AGN (sources with [NeV] or CIV or strong [OIII] lines, i.e. EW([OIII] $\lambda$\,5007) $>$ 3EW(H$\beta$), 1 is a broad line AGN (FWHM line widths greater than 2000\,km\,s$^{-1}$) and the rest are classed as star-forming galaxies (strong Balmer lines and no broad or high-ionization lines) --- see table \ref{table:all} for a summary. An X-ray-detected SPIRE source which satisfies at least one of: log\,[f$_{\rm HX}$/f$_{\rm R}$]$>$-1, HR$>$-0.1 and high-excitation or broad spectral lines, is classed as AGN-dominated in the X-rays and hence not included in our study of the X-ray/IR star-formation correlation. As we shall see below, the combination of these diagnostics enables us to pinpoint all powerful AGN in the sample, as well as identify some less luminous ones. 

Fig. \ref{fig:xray_ir} shows that most hard X-ray detected objects are classed as AGN, whereas due to the higher sensitivity of the soft-band, the soft X-ray detections are an almost even mixture of star-forming galaxies (SFGs) and AGN. In addition, as stated above, it is likely that some of SFGs host AGN not identified by our criteria. There are also 3 sources (objects 35, 43 and 60; see caption of Fig. \ref{fig:xray_ir} and table \ref{table:all}) which are only marginally classified as AGN, because of highly uncertain or unconstrained HR and/or f$_{\rm HX}$/f$_{\rm R}$ values. We incorporate them in the star-forming sub-sample, although our conclusions do not change if we instead include them in the AGN sub-sample.
It is also worth pointing out, that all, bar one, log\,[f$_{\rm HX}$/f$_{\rm R}$]$>$-1 sources have a hard-band X-ray luminosity greater than 10$^{42}$\,erg\,s$^{-1}$.  An X-ray luminosity cut at 10$^{42}$\,erg\,s$^{-1}$ is widely used in extragalactic surveys to separate AGN and star-forming galaxies (e.g. see reviews by Fabbiano 1989\nocite{Fabbiano89}; Brandt $\&$ Hasinger 2005\nocite{BH05}), based on the fact that in the local Universe, where high resolution X-ray studies are possible, the most X-ray luminous starburst systems have X-ray luminosities up to 10$^{42}$\,erg\,s$^{-1}$ in the full band and up to a few times 10$^{41}$\,erg\,s$^{-1}$ in the hard band (see Moran, Halpern $\&$ Helfand 1996, 1999; \nocite{MHH96} \nocite{MLH99} Zezas, Georgantopoulos $\&$ Ward 1998\nocite{ZGW98}). Although, we do not explicitly use a luminosity cut here, so as not to bias our sample against X-ray luminous star-forming sources, we see that its use would produce entirely consistent results with the AGN identified through the f$_{\rm HX}$/f$_{\rm R}$ ratio. 

In total we class 18 objects as AGN-dominated in the X-rays, a fraction of 27 ($\pm$10) per cent. This is higher than the X-ray AGN fraction of the \textit{Spitzer} 70\,$\mu$m-selected samples of Symeonidis et al. (2010\nocite{Symeonidis10}) and Kartaltepe et al. (2010\nocite{Kartaltepe10}) who find only 7 and 9 per cent X-ray-selected AGN respectively. We attribute these differences to (i) the much shallower X-ray surveys that their results are based on and (ii) differences in the relative fractions of LIRGs and ULIRGs in the 3 samples. With respect to the first point, the 2Ms \textit{Chandra} survey that we use here implies a large increase in on-axis sensitivity from a 250\,ks survey, so some of the lower luminosity AGN which are picked up with the 2\,Ms survey would be missed by a shallower X-ray survey. With respect to the second point, our SPIRE sample contains a higher fraction of ULIRGs, which would imply an overall higher AGN fraction, as various observations point towards an increase of the AGN fraction as a function of $L_{\rm IR}$ (e.g. Veilleux et al. 1995\nocite{Veilleux95}; 1999\nocite{VSK99}; Lutz et al. 1998\nocite{Lutz98}; see also Symeonidis et al. 2010; Kartalepe et al. 2010). 

On the other hand, the X-ray AGN fraction of the SPIRE sources is lower than the Alexander et al. (2005b\nocite{Alexander05b}) work on the X-ray properties of SMGs. They find about 75 per cent of their sample hosting X-ray AGN, although as they point out, there is an additional bias due to the radio (1.4\,GHz) selection, resulting in the best estimate of the AGN fraction being closer to 38 per cent (Alexander et al. 2005a\nocite{Alexander05a}). Recent studies by Laird et al. (2010) and Georgantopoulos, Rovilos $\&$ Comastri (2011\nocite{GRC11}) of purely submm-selected SMGs show X-ray AGN fractions of 20-29$\pm$7 per cent and 29$\pm$9 per cent respectively, entirely consistent with our SPIRE AGN fraction.

\section{The soft X-ray/IR star-formation relation}
\label{sec:soft}

\subsection{Trends in the $L_{\rm SX}$/$L_{\rm IR}$ ratio}
\label{sec:ratio}

\begin{figure*}
\centering
\epsfig{file=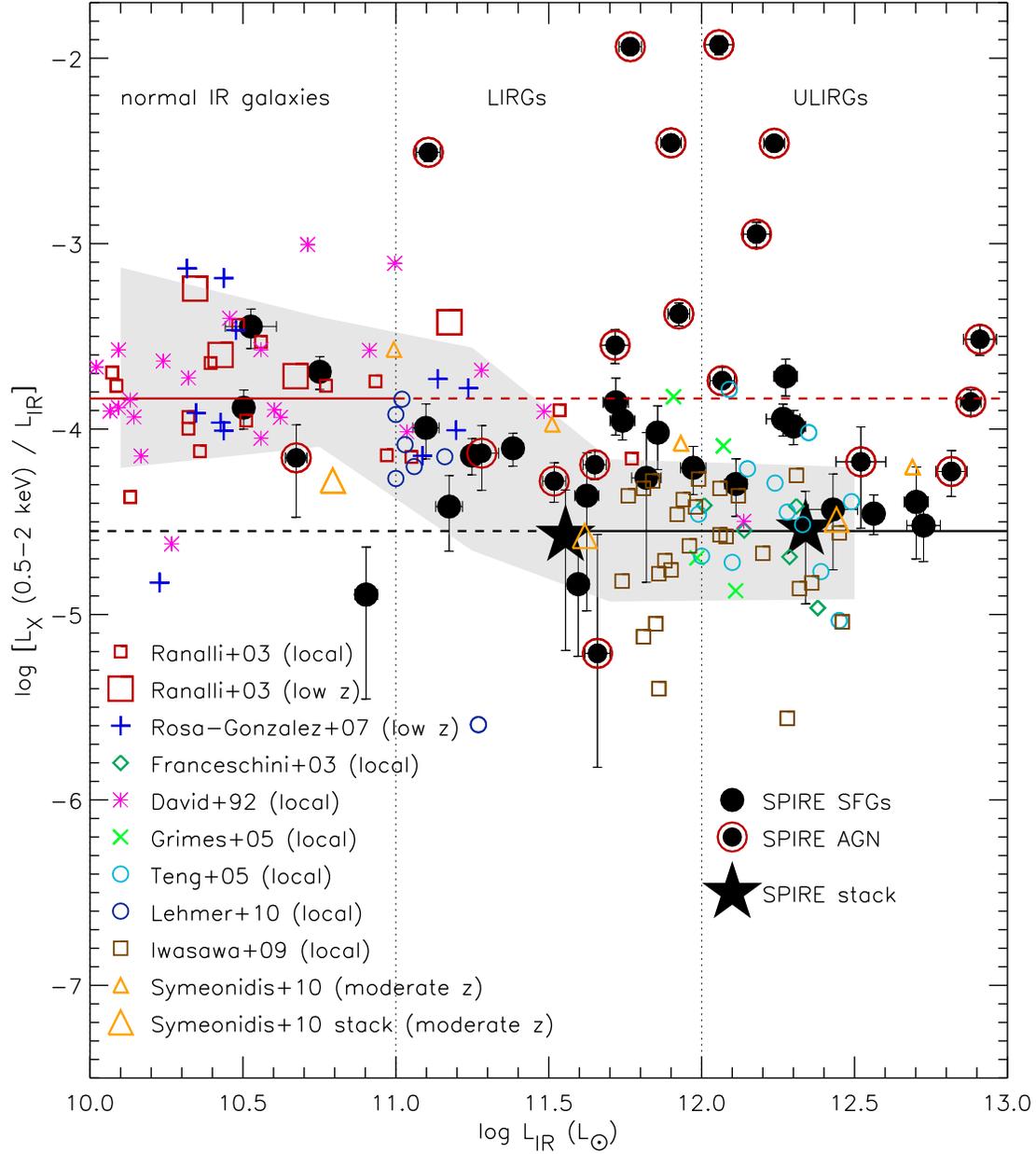,width=0.88\linewidth,clip=}
\caption{ $L_{\rm SX}$/$L_{\rm IR}$ against $L_{\rm IR}$ for the SPIRE SFGs (black filled circles) and AGN (black filled circles with red outline). The large filled stars correspond to the groups of stacked sources (see table \ref{table:stack} for more details). For comparison, we also include various samples from the literature. Pink asterisks: the local \textit{IRAS} 60\,$\mu$m-selected galaxies from David, Jones $\&$ Forman (1992), dark blue open circles: the local LIRG sample of Lehmer et al. (2010; soft X-ray data are courtesy of B. Lehmer), brown open squares:  the local LIRG sample of Iwasawa et al. (2009; 2011), light blue open circles: local/low-redshift ULIRGs from Teng et al. (2005), green open diamonds: local ULIRGs from Franceschini et al. (2003), light green crosses: local ULIRGs from Grimes et al. (2005), small red open squares: local optically identified HII galaxies from Ranalli, Comastri $\&$ Setti (2003), large red open squares: low redshift X-ray/radio identified star-forming galaxies from Ranalli, Comastri $\&$ Setti (2003), dark blue plus signs: low redshift X-ray selected CDFS sources from Rosa-Gonz{\'a}lez et al. (2007) and orange triangles (small for detections and large for stacking): moderate redshift ($\left\langle z \right\rangle$=0.6), 70\,$\mu$m-selected galaxies from Symeonidis et al. (2010). The grey shaded region correponds to the smoothed mean and 1$\sigma$ envelopes of the local and low redshift samples calculated in 5 luminosity bins (10$<$\,log\,($L_{\rm IR}$/L$_{\odot}$)\,$<$12.5 in steps of 0.5 dex). The vertical dotted lines separate the 3 IR luminosity classes that the sources fall into. The black horizontal line is equation \ref{equation1} (solid part), extrapolated to lower luminosities than were used to derive it (dashed part). The red horizontal line corresponds to the Ranalli et al. (2003) linear soft X-ray/IR correlation (solid part; see Section \ref{sec:ratio}) extrapolated to higher luminosities than were used to derive it (dashed part).}
\label{fig:ratio_soft}
\end{figure*}

In order to examine the X-ray/IR properties of the SPIRE SFGs, we examine the $L_{\rm SX}$/$L_{\rm IR}$ ratio as a function of $L_{\rm IR}$ in Fig. \ref{fig:ratio_soft}. For completeness, we also show the sources hosting AGN, clearly marked on the plot. However we do not include them in our analysis of the X-ray/IR relation, focusing solely on star-forming galaxies.
Note that, due to small number statistics, any kind of evolution in
the $L_{\rm SX}$/$L_{\rm IR}$ ratio cannot be measured within the
SPIRE SFGs. Nevertheless, we opt to compare their properties with
other samples taken from the literature. These are the local
\textit{IRAS} 60\,$\mu$m-selected galaxies from David, Jones $\&$
Forman (1992), the local LIRG sample from Lehmer et al. (2010; soft
X-ray data are courtesy of B. Lehmer),  the local LIRG sample from
Iwasawa et al. (2009; 2011\nocite{Iwasawa09}\nocite{Iwasawa11}), local/low-redshift ULIRGs from Teng et al. (2005\nocite{Teng05}), local ULIRGs from Franceschini et al. (2003), local ULIRGs from Grimes et al. (2005\nocite{Grimes05}),
local optically identified HII galaxies and low redshift X-ray/radio identified star-forming galaxies from Ranalli, Comastri $\&$ Setti (2003), low redshift X-ray selected CDFS sources from
Rosa-Gonz{\'a}lez et al. (2007) and moderate redshift ($\left\langle z \right\rangle$=0.6), 70\,$\mu$m-selected galaxies from Symeonidis et
al. (2010). In order to be consistent, we base our comparison on objects from these samples, which do not host identifiable AGN (AGN
are identified by various methods in these studies) and with a far-IR (40--120\,$\mu$m) or total IR (8--1000\,$\mu$m) luminosity in the
10$<$\,log\,($L_{\rm IR}$/L$_{\odot}$)\,$<$13 range.  Where required, we make appropriate conversions to the cosmology we use in this paper,
as well as to the soft X-ray band range of 0.5--2\,keV. Wherever necessary we also convert far-IR luminosity ($L_{\rm FIR}$) to total
infrared luminosity ($L_{\rm IR}$) assuming $L_{\rm FIR}$ is 70 per cent of $L_{\rm IR}$ (Helou et al. 1988). All X-ray luminosities are
in the 0.5-2\,keV band, corrected only for Galactic absorption and \textit{K}-corrected (with $\Gamma$=1.9). 

In Fig. \ref{fig:ratio_soft}, we also plot the linear relation from Ranalli, Comastri $\&$ Setti (2003), as above appropriately modified by converting far-IR luminosity ($L_{\rm FIR}$) to total infrared luminosity ($L_{\rm IR}$). Note that the Ranalli et al. relation was derived predominantly with a $L_{\rm IR}$\,$<$10$^{11}$\,L$_{\odot}$ sample, a large fraction of which were normal star-forming galaxies ($L_{\rm IR}$\,$<$10$^{10}$\,L$_{\odot}$)  and is not necessarily representative of higher luminosity sources. However for comparison pusposes we extrapolate it up to $L_{\rm IR}$\,=10$^{13}$\,L$_{\odot}$. Although Ranalli et al. discuss that their sample is consistent with both $L_{\rm SX}$\,$\propto$\,$L_{\rm IR}$ and $L_{\rm SX}$\,$\propto$\,$L_{\rm IR}^{0.87}$, our subsequent comparison uses the linear relationship as a point of reference, as it describes a scenario where X-ray emission scales with the star-formation rate. 
The linear Ranalli et al. relation derived for local galaxies is also seen to hold at low ($z <$0.4) redshifts for sources of equivalent luminosity, e.g. see Rosa-Gonz{\'a}lez et al. (2007)\nocite{Rosa-Gonzalez07}.

Apart from the Symeonidis et al. (2010) sources, all other comparison samples are at $z<$0.4, with the majority at $z<$0.2. In order to quantify any change in the $L_{\rm SX}$/$L_{\rm IR}$ ratio as a function of IR luminosity, we group them in 5 infrared luminosity bins (10$<$\,log\,($L_{\rm IR}$/L$_{\odot}$)\,$<$12.5 in steps of 0.5 dex), calculating the mean $L_{\rm SX}$/$L_{\rm IR}$ ratio and 1$\sigma$ envelopes for each bin, shown as a smooth grey shaded region in Fig. \ref{fig:ratio_soft}. Although the scatter is large, the average $L_{\rm SX}$/$L_{\rm IR}$ ratio of these 9 $z<$0.4 samples is seen to decrease with increasing IR luminosity: for log\,($L_{\rm IR}$/L$_{\odot}$)\,$\gtrsim$11.5 sources, the mean ratio is 0.7 dex lower than for normal IR galaxies. The linear Ranalli et al. relation (red line in Fig. \ref{fig:ratio_soft}), which predicts a constant $L_{\rm SX}$/$L_{\rm IR}$ ratio of -3.84, overpredicts the properties of the more $L_{\rm IR}$-luminous sources. The $L_{\rm SX}$/$L_{\rm IR}$ ratio for the latter drops to an average value of -4.55$\pm$0.05 (1$\sigma$ scatter of 0.37). Taking the soft X-ray emission to be proportional to $L_{\rm IR}$ (and hence the star-formation rate), the $L_{\rm SX}$/$L_{\rm IR}$ relation for local $L_{\rm IR}$\,$\gtrsim$3\,$\times$10$^{11}$\,L$_{\odot}$ galaxies can thus be written as:
\begin{equation}
 \rm log\,\textit{L}_{\rm SX}=\rm log\,\textit{L}_{\rm IR}  - 4.55
\label{equation1}
\end{equation}

The result that more $L_{\rm IR}$-luminous sources are soft X-ray deficient with respect to their IR emission compared to their less luminous counterparts has been noted before by Franceschini et al. (2003) and Grimes et al. (2005). Grimes et al. (2005) discuss that although the X-ray spectra of both normal IR galaxies and ULIRGs are dominated by the thermal component at low energies $<$\,2\,keV, the latter have a much larger characteristic size of their diffuse X-ray emitting regions; the radius which encloses 90 per cent of the flux extends up to 30\,kpc in ULIRGs. However, the discrepancies in soft-band X-ray emission cannot be reconciled by missing flux, as Grimes et al. show that, once bright point source flux is removed, the surface brightness at the 90 per cent radius is roughly constant in both galaxy types. Instead, the observation that the diffuse X-ray emission in ULIRGs extends to much larger scales suggests that the density and hence emissivity of the X-ray emitting gas might be lower, which could in part be responsible for making high luminosity sources soft-X-ray-deficient.

The low redshift samples we use here are highly complete, with the X-ray association rate greater than 80 per cent, higher than the association rate of 51 per cent for the SPIRE SFGs (out of the 49 sources in our sample classified as SFGs, only 25 are X-ray detected). This implies that for the comparison between these low redshift samples and the SPIRE SFGs, it is appropriate to include the stacked results for the latter. Fig. \ref{fig:ratio_soft} shows that the SPIRE SFGs have ratios consistent with the scatter seen in their local equivalents, within 2$\sigma$. The average $L_{\rm SX}$/$L_{\rm IR}$ ratio for LIRGs and ULIRGs in the SPIRE sample is estimated using the detections and average stacked luminosities, the latter weighted by the number of sources in the stacked groups. We find $\left\langle L_{\rm SX}/L_{\rm IR} \right\rangle$= -4.43 with a 1$\sigma$ scatter of 0.37, consistent with the average ratio of their local counterparts (equation \ref{equation1}). Similarly, the normal IR galaxies ($L_{\rm IR}$\,$<$10$^{11}$\,L$_{\odot}$) in the SPIRE sample appear to have properties consistent with those of local normal IR galaxies, described by the Ranalli et al. relation, although due to small number statistics this result is uncertain. The SPIRE AGN display large scatter in log\,[$L_{\rm SX}$/$L_{\rm IR}$] and many clearly stand out with values $>$-4. However, some show ratios consistent with the SPIRE SFGs, either because their soft X-ray emission is intrinsically low or because of substantial absorption in the soft X-rays. Equivalently, some of the SFG-classified galaxies could in fact be AGN-dominated in the X-rays. Inevitably, separating the AGN and SB contribution in these sources may only be possible with high resolution X-ray spectra, outside the capabilities of current X-ray observatories. 

Although the comparison between local and high redshift SFGs is fairly straight forward, an issue which needs to be addressed is that at z$>$0, the soft band starts receiving emission from higher energy X-rays. After $\sim$2\,keV (rest-frame), the contribution from diffuse gas declines whereas that of point sources increases, with the X-ray spectrum changing from predominantly thermal to predominantly power-law (e.g. Persic $\&$ Rephaeli 2002; Franceschini et al. 2003; Grimes et al. 2005). For most of the SPIRE SFGs this is not a serious problem since at $z<$1 the X-rays received by the soft band are still largely of thermal origin. On the other hand, for the higher redshift ULIRGs, the rest-frame 0.5--2\,keV part of the X-ray spectrum is mostly unconstrained by our data. 
The \textit{K}-correction with $\Gamma$=1.9 would be sufficient to correct for this, under the assumption that the X-ray SEDs of high redshift SFGs are similar to those of their local counterparts and hence relatively flat (in $\nu$f$_{\nu}$). This assumption, i.e. that the ratio of thermal to non-thermal emission stays relatively constant with redshift, is reasonable for two reasons: (i) $z<$1 SPIRE SFGs have $L_{\rm SX}$/$L_{\rm IR}$ ratios consistent with their local counterparts, so it would take fundamental changes in the origin of the soft X-ray emission for the $z>$1 ULIRGs in our sample to be different to $z<$1 LIRGs and ULIRGs; (ii) in the hard X-rays, where the 5--10\,keV band receives emission from the non-thermal part of the spectrum at all redshifts, a simple \textit{K}-correction is sufficient to rest-frame the flux and we find consistent properties between local and high redshift sources (see section \ref{sec:hard}). To summarise: given that our data shows no evolution in the $L_{\rm SX}$/$L_{\rm IR}$ ratio up to $z<$1 and no evolution in the $L_{\rm HX}$/$L_{\rm IR}$ ratio up to $z<$2, it is reasonable to assume that the $L_{\rm SX}$/$L_{\rm IR}$ ratio is also constant up to $z<$2, although this would only be formally settled with the availability of $<$0.5\,keV X-ray data.

\subsection{The $L_{\rm SX}$/$L_{\rm IR}$ ratio as a tracer of stellar evolution}
\label{sec:models}

\begin{figure}
\centering
\epsfig{file=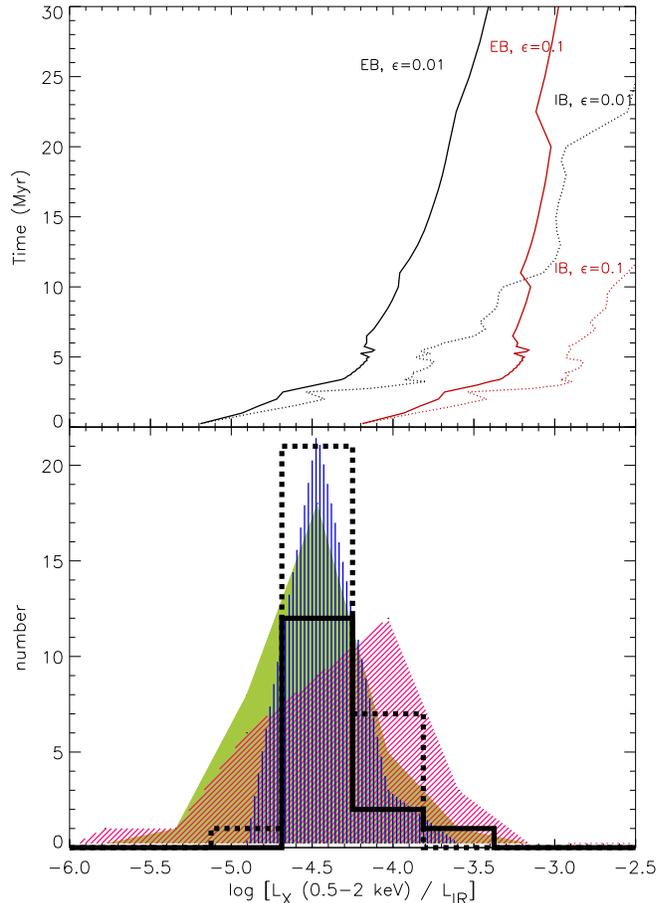,width=8.5cm}
\caption{ In the top panel, we plot the Mas-Hesse, Oti-Floranes $\&$ Cervi{\~n}o (2008) models which use the $L_{\rm SX}$/$L_{\rm IR}$ ratio as a tracer of stellar evolution and compute its variation over 30\,Myr, based on the evolutionary synthesis models of Cervi{\~n}o, Mas-Hesse $\&$ Kunth (2002), assuming solar metallicity and a Salpeter IMF with stellar mass limits of 2--120\,M$_{\odot}$. EB stands for extended bursts, IB stands for instantaneous burst. $\epsilon$ is the efficiency by which mechanical energy is used to heat up the X-ray emitting gas. In the lower panel we show the distribution in log\,[$L_{\rm SX}$/$L_{\rm IR}$] ratio for the SPIRE LIRGs and ULIRGs (dashed and solid black histograms respectively) compared to equivalent sources from the samples described in section \ref{sec:ratio}. These are local/low redshift LIRGs (pink region) and ULIRGs (green filled region) and the moderate redshift log\,($L_{\rm IR}$/L$_{\odot}$)\,$>$11 sources from Symeonidis et al. (2010; blue region). Note that for clarity, the comparison samples are not plotted as histograms, but rather we interpolate between the total number of sources in each log\,[$L_{\rm SX}$/$L_{\rm IR}$] bin and shade the area underneath. Also, in the case of stacked data, each source in the stacked group is counted individually in the relevant bin. The log\,[$L_{\rm SX}$/$L_{\rm IR}$] bins are 0.44\,dex in width.}
\label{fig:ratio_soft_wmodels}
\end{figure}

Mas-Hesse, Oti-Floranes $\&$ Cervi{\~n}o (2008, hereafter MOC08) use the $L_{\rm SX}$/$L_{\rm IR}$ ratio as a tracer of stellar evolution and compute its variation over 30\,Myr, based on the evolutionary synthesis models of Cervi{\~n}o, Mas-Hesse $\&$ Kunth (2002), which assume solar metallicity and a Salpeter IMF with stellar mass limits of 2--120\,M$_{\odot}$. 
In Fig. \ref{fig:ratio_soft_wmodels} we plot the MOC08 models (top panel) and the distribution in $L_{\rm SX}$/$L_{\rm IR}$ for the LIRGs and ULIRGs in the samples described in section \ref{sec:ratio} (lower panel). 
According to MOC08, the $L_{\rm SX}$/$L_{\rm IR}$ ratio is highly
sensitive to the star-formation history and the efficiency by which
energy is coupled to the ISM and as a result it can span an order of
magnitude at any given epoch. The MOC08 models shown here were derived
for two different star-formation modes: stars forming during an
instanteneous burst (IB) or an extend period of time (EB), although
realistically, EB systems are not thought to be younger than $\sim$10
Myr, as they are supposed to be a composite of many instantaneous
bursts mimicing constant star formation. The MOC08 models are also
shown as functions of $\epsilon$ (=0.01 or 0.1), which is the fraction
of energy incident onto the ISM that heats the X-ray emitting
gas. Note that MOC08 neglect point sources, as these are estimated to contribute a negligible amount (up to 15 per cent) to the soft X-ray emission in the early stages of the starburst evolution examined by their models.

MOC08 discuss that after a few Myrs, IR emission decreases sharply in the IB mode as the most massive stars reach the end of their lifetimes and therefore the intensity by which the dust is being heated significantly decreases. On the other hand in the EB mode, IR emission stays relatively constant as massive stars are being continuously replenished. The soft X-ray luminosity undergoes a rapid increase during the first few Myrs, due to the injection of energy into the ISM from the winds of the most massive stars in both IB and EB modes. Cervi{\~n}o, Mas-Hesse $\&$ Kunth (2002) propose that, as a result, X-ray emission is strongly metallicity dependent during the first few Myr. Subsequently, injection of energy is dominated by supernovae so the soft X-ray luminosity increases for EB models, whereas it levels off in IB models. When translated to a $L_{\rm SX}$/$L_{\rm IR}$ ratio, this implies a continuous increase in the ratio for IB models and a smaller rate of increase in the EB scenario. In addition, for a given efficiency and at any given epoch, the IB mode of star formation produces an X-ray excess compared to the EB mode, because of the larger number of supernovae that heat up the gas. 

In the lower panel of Fig. \ref{fig:ratio_soft_wmodels}, we show the distribution in log\,[$L_{\rm SX}$/$L_{\rm IR}$] for the SPIRE LIRGs and ULIRGs compared to the samples presented in section \ref{sec:ratio} grouped into local/low redshift LIRGs, local/low redshift ULIRGs and moderate redshift log\,($L_{\rm IR}$/L$_{\odot}$)\,$>$11 sources from Symeonidis et al. (2010). The log\,[$L_{\rm SX}$/$L_{\rm IR}$] bins are 0.44\,dex in width and in the case of stacked data, each source in the stacked group is counted individually in the relevant luminosity bin. 

As described in section \ref{sec:ratio} and also clearly seen here, there is significant overlap between the 5 groups, indicative of little or no change in the $L_{\rm SX}$/$L_{\rm IR}$ ratio of LIRGs and ULIRGs with redshift. However, note that stellar evolution might not be the only relevant factor in shaping the X-ray properties of SFGs. As discussed in Section \ref{sec:ratio}, the characteristic size of the X-emitting region could influence the strength of X-ray emission and might explain why there is a group of local LIRGs with higher $L_{\rm SX}$/$L_{\rm IR}$ ratios than the bulk of the other objects (far-right section of the pink shaded region in the lower panel of Fig. \ref{fig:ratio_soft_wmodels}) --- these are the lowest luminosity LIRGs in the sample ($L_{\rm IR}$\,$\sim$\,10$^{11}$\,L$_{\odot}$; see also Fig. \ref{fig:ratio_soft}).

\begin{figure*}
\epsfig{file=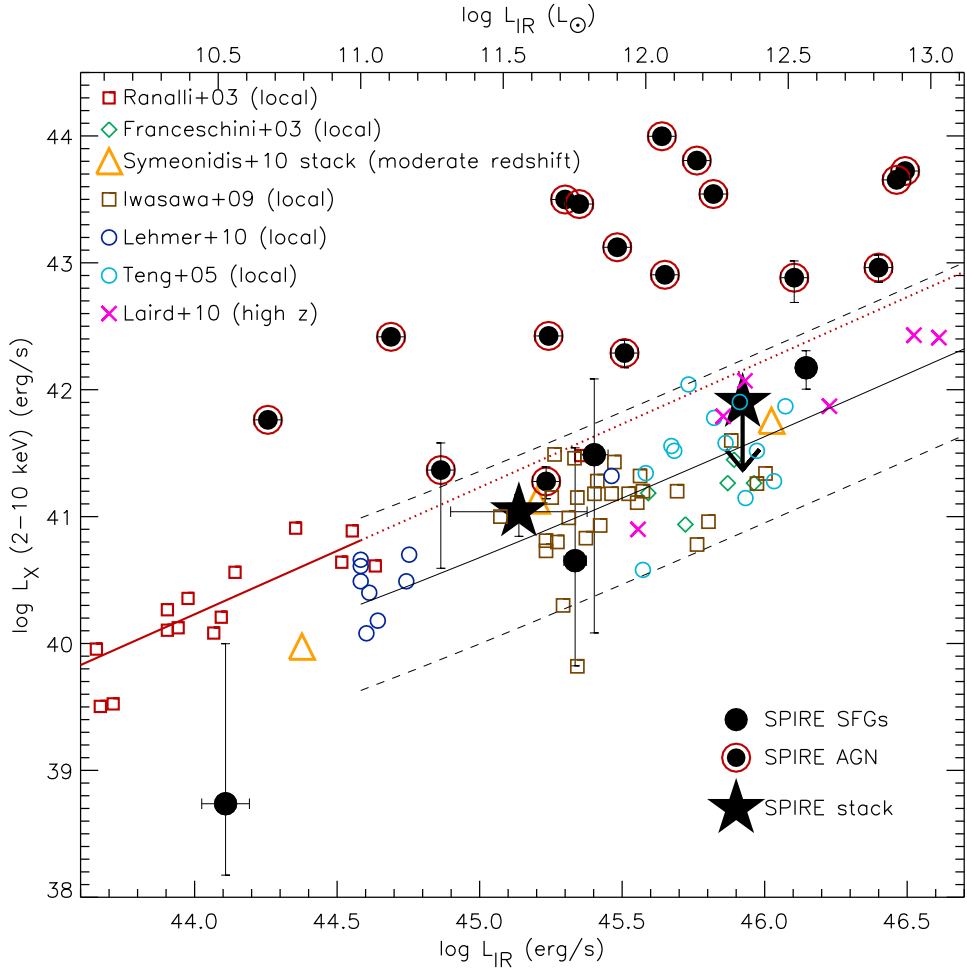,width=14cm}
\caption{Hard X-ray luminosity ($L_{\rm HX}$) versus $L_{\rm IR}$ for the SPIRE SFGs (black filled circles) and AGN (black filled circles with red outline); large filled stars correspond to the stacking groups (see table \ref{table:stack}). Note that only LIRG group has a stacked signal in the hard X-rays, whereas for the ULIRGs we plot the 3\,$\sigma$ upper limit. For comparison we show the Ranalli et al. $L_{\rm HX}$-$L_{\rm IR}$ linear relation (red solid line) extrapolated to higher luminosities than were used to derive it (dashed red line). Some of the samples shown in Fig. \ref{fig:ratio_soft} are also included --- the Ranalli et al. (2003) star-forming galaxies (red empty squares), the Franceschini et al. (2003) ULIRGs (green diamonds), the Symeonidis et al. (2010) sample (orange triangles, consisting of stacked data only in the hard X-rays), the local LIRG/ULIRG samples from Iwasawa et al. (2009, brown squares) and Lehmer et al. (2010, blue circles) and the $z$\,$<$\,0.16 ULIRG sample from Teng et al. (2005, blue circles). Also included is the submm galaxy (SMG) sample of Laird et al. (2010; pink crosses). The solid black line is the Lehmer et al. (2010) equation (equation \ref{equation_L4} in Section \ref{sec:hard}; equivalent to equation \ref{equation_L5} for $L_{\rm IR}$\,$\gtrsim$10$^{11}$\,L$_{\odot}$) and 2\,$\sigma$ scatter of 0.68\,dex (dashed black lines).}
\label{fig:sfr_hard}
\end{figure*}

Fig. \ref{fig:ratio_soft_wmodels} illustrates that the X-ray luminosities of the majority of objects are sufficiently low for the average log\,[$L_{\rm SX}$/$L_{\rm IR}$] to be $\lesssim$-4 and according to the MOC08 models, consistent with very young ages ($\lesssim$\,10\,Myr) and low efficiency ($\epsilon$=0.01). For the few remaining objects with log\,$[L_{\rm SX}/\textit L_{\rm IR}] \gtrsim$\,-4, age, efficiency and mode of star-formation are degenerate, but apart from the EB, $\epsilon$\,=0.01 model, they are also compatible with young ages up to 10\,Myr. Note that during the very early stages ($\lesssim$5\,Myr) of star-formation, the $L_{\rm SX}$/$L_{\rm IR}$ ratio is sensitive to the IMF (slope, mass limits) --- see Ot{\'{\i}}-Floranes $\&$ Mas-Hesse (2010\nocite{OM10}). For example, changing the IMF mass limits  from 2--120\,M$_{\odot}$ in the MOC08 models to 0.1--100\,M$_{\odot}$ (e.g. Kennicutt 1998), can reduce the $L_{\rm SX}$/$L_{\rm IR}$ ratio by up to 30 per cent, whereas assuming a top-heavy IMF (slope= -1) would yield higher $L_{\rm SX}$/$L_{\rm IR}$ values in the IB scenario. Nevertheless, despite any changes in the IMF and/or efficiency (which cannot be quantified currently), the X-ray/IR ratio of LIRGs and ULIRGs is overall consistent with very young stellar ages. This might seem stange at first, as LIRGs and ULIRGs have high stellar masses and therefore a substantial older population. The fact that these systems are represented by young ages in the MOC08 models suggests that both the soft X-ray and infrared emission must be dominated by the current star-formation episode, with minimal contribution from older stellar populations or regions of lower SFR. This is not surprising as the main factors in heating the X-ray emitting gas are supernovae and stellar winds, both of which are characteristic of massive stars, implying that massive stellar evolution is the key player in the soft X-ray output of a high SFR galaxy. Moreover, although total infrared luminosity traces the starburst episode in LIRGs and ULIRGs up to $\sim$100\,Myr (e.g. see review by Kennicutt 1998 and references therein), many models (e.g. Devereux $\&$ Young 1990\nocite{DY90}; Mas-Hesse $\&$ Kunth 1991\nocite{MHK91}; Cervino $\&$ Mas-Hesse 1994\nocite{CMH94}; Devereux $\&$ Hameed 1997\nocite{DH97}) support that infrared luminosity peaks at $\lesssim$\,10\,Myr, suggesting that the peaks of IR and X-ray emission are roughly coincident for galaxies undergoing starburst episodes.

\section{The hard X-ray/IR star-formation correlation}
\label{sec:hard}

\begin{figure}
\epsfig{file=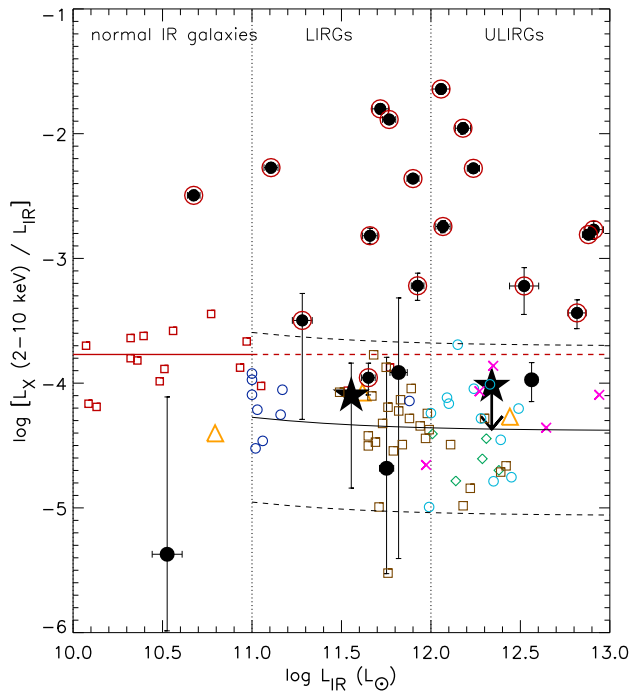,width=9cm}
\caption{The $L_{\rm HX}$/$L_{\rm IR}$ ratio plotted against $L_{\rm IR}$ for the SPIRE SFGs (black filled circles) and AGN (black filled circles with red outline); large filled stars correspond to the stacking groups (see table \ref{table:stack}). The vertical dotted lines separate the 3 IR luminosity classes that the sources fall into. All other symbols and lines are the same as in Fig. \ref{fig:sfr_hard}.}
\label{fig:ratio_hard}
\end{figure}

Fig. \ref{fig:sfr_hard} shows hard X-ray luminosity ($L_{\rm HX}$) versus $L_{\rm IR}$ for the SPIRE sample. This is also plotted as a ratio in Fig. \ref{fig:ratio_hard} to facilitate comparison with Fig. \ref{fig:ratio_soft}. We also plot the Ranalli et al. $L_{\rm HX}$-$L_{\rm IR}$ linear relation, again appropriately modified to convert $L_{\rm FIR}$ to $L_{\rm IR}$ and extrapolated to higher luminosities than were used to derive it (see section \ref{sec:soft}). 
Some of the samples shown in Fig. \ref{fig:ratio_soft} are also
included here --- the Ranalli et al. (2003) star-forming galaxies, the
Franceschini et al. (2003) ULIRGs, the Symeonidis et al. (2010) sample
(consisting of stacked data only in the hard X-rays), the local
LIRG/ULIRG samples from Iwasawa et al. (2009\nocite{Iwasawa09}) and
Lehmer et al. (2010\nocite{Lehmer10}) and the $z$\,$<$\,0.16 ULIRG
sample from Teng et al. (2005\nocite{Teng05}). Also included is the
submm galaxy (SMG) sample of Laird et al. (2010\nocite{Laird10}). All
comparison samples comprise sources which do not host AGN according
the classifications given by the authors, however note that as for
the soft X-rays, there is always
a non-negligible possibility that some have been misclassified. Finally, all necessary conversions were performed to the data in order for the comparison to be consistent (see Section \ref{sec:soft} for an account). 

Figs. \ref{fig:sfr_hard} and \ref{fig:ratio_hard} show that local/low redshift LIRGs/ULIRGs have significantly lower hard X-ray luminosities than what is predicted by the Ranalli et al. (2003) relation derived for normal star-forming galaxies, a result that has been well established by various studies e.g. Persic $\&$ Rephaeli (2002), Franceschini et al. (2003), Persic et al. (2004), Iwasawa et al. (2009), Lehmer et al. (2010). This is consistent with the picture seen in the soft X-rays (Fig. \ref{fig:ratio_soft}), where the discrepancies are thought to be primarily due to differences in the extent of X-ray emitting region, the efficiency of energy coupling to the ISM and mode of star-formation. With respect to the hard X-rays, Persic $\&$ Rephaeli (2002), Franceschini et al. (2003) and Persic et al. (2004) argue that the differences in X-ray emission could arise because of varying contributions from HMXBs and LMXBs. In LIRGs and ULIRGs, emission from HMXBs dominates the hard X-ray luminosity, but in lower luminosity star-forming galaxies, LMXBs also contribute substantially, implying that for such sources, hard X-rays are not a reliable tracer of star-formation. Lehmer et al. (2010), derive an expression for the hard X-ray luminosity as a function of both the star-formation rate (SFR, i.e. emission from HMXBs) and the stellar mass (M, i.e. emission LMXBs), 
\begin{equation}
 L_{\rm HX}=\alpha M+\beta SFR.
\label{equation_L1}
\end{equation}
They point out that in the low SFR regime (SFR$<$5\,$M_{\odot}\,\rm
yr^{-1}$), there seems to be an intrisic correlation between SFR and
stellar mass. This causes the first term in equation \ref{equation_L1}
to dominate, resulting in the observed linear relation between $L_{\rm
  HX}$ and SFR (e.g. in the Ranalli et al. 2003 sample), even though
the former quantity is not directly related to the latter; this is also shown
in Grimm, Gilfanov $\&$ Sunyaev (2003); Gilfanov, Grimm $\&$ Sunyaev
(2004); Colbert et al. (2004)\nocite{Colbert04}. 
In the high SFR regime  (SFR$>$5\,$M_{\odot}$\,yr$^{-1}$), Lehmer et al. show that stellar mass only weakly correlates with the star-formation rate [log\,(\textit{M})\,=\,10.4+0.3\, log\,(\textit{SFR})], so the relationship between $L_{\rm HX}$ and $L_{\rm IR}$ becomes 
\begin{equation}
\rm \textit{L}_{\rm HX}=(1.9\times10^{26})\textit{L}_{\rm IR}^{0.3}+(4.15\times10^{-5})\textit{L}_{\rm IR} 
\label{equation_L4}
\end{equation}
where $\rm SFR (M_{\odot}\rm yr^{-1})\,=\,9.8 \times 10^{-11}\,L_{\rm IR}$ (Kenniccutt et al. 1998), $\alpha$=9.05$\times$10$^{28}$\,erg\,s$^{-1}$\,M$_{\odot}^{-1}$ and $\beta$=1.62$\times$10$^{39}$\,erg\,s$^{-1}$\,(M$_{\odot}\,\rm yr^{-1}$)$^{-1}$ and both luminosities are in erg\,s$^{-1}$. This is plotted in Figs. \ref{fig:sfr_hard} and \ref{fig:ratio_hard} together with the 2$\sigma$ scatter of 0.68 dex quoted in Lehmer et al. (2010). Note that for high SFRs and $L_{\rm IR}$\,$\gtrsim$10$^{11}$\,L$_{\odot}$ a simpler version of this equation is derived by dropping the first term in equation \ref{equation_L1}, whereby $L_{\rm HX}$ becomes proportional to $L_{\rm IR}$:
\begin{equation}
 \rm log\,\textit{L}_{\rm HX}=log\,\textit{L}_{\rm IR}  - 4.38 
\label{equation_L5}
\end{equation} 
This is about 0.6 dex lower than the relation of Ranalli et al. (2003) derived for normal star-forming galaxies.

Note that due to the lower sensitivity of \textit{Chandra} in the hard X-rays (almost an order of magnitude lower than in the soft X-rays), we only detect 4 SPIRE SFGs, albeit at low signal to noise, 3 of which are ambiguous and could potentially be AGN dominated (see Section \ref{sec:agn_fraction}). Nevertheless, our overall results, which include the stacking, are in agreement with the Lehmer et al. relation  (equation \ref{equation_L5}) derived for local LIRGs and ULIRGs within the 2$\sigma$ scatter. This is also the case for the moderate and high redshift samples of Symeonidis et al. (2010) and Laird et al. (2010), on par with a no-evolution scenario for the $L_{\rm HX}$/$L_{\rm IR}$ ratio of SB-dominated LIRGs and ULIRGs up to z$\sim$3. 
In terms of the SPIRE AGN, we see that most occupy a different parameter space to the star-forming galaxies. This suggests that in the absense of hardness ratio, R-band magnitude or optical line measurements, the hard X-ray/IR correlation can successfully be used to separate AGN from star-forming galaxies.

\section{Summary and Conclusions}
\label{sec:conclusions}

Our analysis is based on a sample of 67, spectroscopically-identified, SPIRE-detected, submm-bright ($f_{250 \rm \mu m}$\,$>$17.4\,mJy; $>$3$\sigma$ including confusion noise) galaxies in the GOODS-N field, observed as part of the \textit{Herschel}/HerMES survey, the majority of which are LIRGs and ULIRGs at z\,$\sim$1. Taking advantage of the 2\,Ms \textit{Chandra} Deep Field North survey, we investigate the SPIRE sources' soft and hard X-ray emission relative to their infrared output and compare our results to several other samples from the literature. The focus of our study is the X-ray/IR correlation for star-forming IR-luminous galaxies at z$\sim$1. Our sample is selected to be submm-bright, implying that all objects are star-formation-dominated in the infrared, thus we identify AGN through X-ray diagnostics and optical spectral lines. We class as AGN-dominated in the X-rays, objects which satisfy at least one of the following criteria: a hard X-ray/R-mag flux ratio (log\,[f$_{\rm HX}$/f$_{\rm R}$]) greater than -1, a hardness ratio (HR) greater than -0.1 and the presence of high-excitation or broad optical lines. Out of the 43 X-ray detected sources, 18 are identified as AGN, amounting to a fraction of 27 ($\pm$10) per cent in the SPIRE sample. Note that our aim is to exclude any obvious AGN, but we cannot ensure a pure star-forming sample, since low luminosity and/or unabsorbed AGN will be missed. 
Once obvious AGN are eliminated, our analysis is based on 24 X-ray detected SPIRE objects and 23 non-X-ray detected SPIRE sources, whose X-ray properties we assess through stacking. 

We conclude that the X-ray/IR properties of star-formation-dominated SPIRE galaxies, quantified by the soft and hard X-ray/IR luminosity ratios ($L_{\rm SX}$/$L_{\rm IR}$ and $L_{\rm HX}$/$L_{\rm IR}$), are consistent with those observed in the local ($z<0.1$) Universe. They are also consistent with the ratios of equivalent sources at more moderate ($z\sim0.6$) and higher redshifts ($z\sim2$), indicating that, overall, there is no evidence for evolution of the X-ray/IR correlation for IR-luminous galaxies with redshift. Note that, although none of the samples we present here are entirely impervious to AGN contamination, the main outcome of our study is robust, in the sense that once all identifiable bona-fide AGN are excluded, the remaining sources have consistent average X-ray/IR ratios at all redshifts. 

The linear nature of the X-ray/IR correlation describes a scenario whereby both X-ray and IR emission trace the star-formation rate and is thus important in calibrating X-ray emission as a star formation indicator, providing an alternative census of the global SFR density. This will be particularly applicable in future deeper X-ray surveys, where the X-ray number counts at the faint flux end will be dominated by star-forming galaxies.

\onecolumn
\begin{table}
\centering
 \caption{The full SPIRE sample of 84 SPIRE-detected, submm-bright ($f_{250 \rm \mu m}$\,$>$17.4\,mJy) galaxies in the GOODS-N field, observed as part of the \textit{Herschel}/HerMES survey. The columns are as follows:
(1) ID, (2) (3) IRAC-3.6\,$\mu$m RA and DEC, (4) spectroscopic redshift (available only for 67 sources), (5) total infrared luminosity (8--1000\,$\mu$m), (6) the cross-matched ID from
the IC catalogue of HDFN Chandra data, (7) (8) (9) (10) the flux and luminosity in the
soft and hard bands (SX stands for `soft X-ray band' and HX stands for
`hard X-ray band'), (11) the hardness ratio (HR), (12) the $f_{\rm HX}$/$f_{\rm R}$ ratio used to identify AGN (see section \ref{sec:agn_fraction}), computed for sources which are detected in both the R-band and the hard X-rays. Note that, although we provide the corresponding HDFN X-ray ID from the IC catalogue, the fluxes and upper limits we present here are all derived using the IRAC-3.6\,$\mu$m position of each SPIRE source to extract counts from the X-ray map down to a Poisson significance $<3\times10^{-3}$ --- see Section \ref{sec:xraydata} for details. `$<$\,f$_{\rm X}$' corresponds to the 3$\sigma$ upper limit. Column (13) refers to the classification of the
source by optical spectroscopy (Barger et al. 2005; Trouille et al.
2008). SFG stands for `Star Forming Galaxy', sources with strong
Balmer lines and no broad or high-ionization lines. HEL AGN stands for
`High Excitation Line AGN', sources with [NeV] or CIV lines or strong
[OIII] lines, i.e. EW([OIII] $\lambda$\,5007) $>$ 3EW(H$\beta$). BL AGN stands for `Broad-line AGN', sources with 
FWHM line widths greater than 2000\,km\,s$^{-1}$. Note that the Trouille et al. (2008) classification applies only to X-ray sources. 
Where there is no information available, the spectrum could not be classified. Column (14) refers to whether the source
is SB or AGN dominated in the X-rays, according to the criteria
outlined in section \ref{sec:agn_fraction}. The 3 ambiguous sources (labelled `Amb'), are included in the SFG sub-sample, but our conclusions do not change if instead they are included in the AGN sub-sample. }
\begin{tabular}{|@{\hspace{-0.1mm}}l||@{\hspace{-0.1mm}}l||@{\hspace{-0.1mm}}l||@{\hspace{-0.1mm}}l||@{\hspace{-0.1mm}}l||@{\hspace{-0.1mm}}l||@{\hspace{-0.1mm}}l||@{\hspace{-0.1mm}}l||@{\hspace{-0.1mm}}l||@{\hspace{-0.1mm}}l||@{\hspace{-0.1mm}}l||@{\hspace{-0.1mm}}l||@{\hspace{-0.1mm}}l||@{\hspace{-0.1mm}}l|}

 \hline
 \hline
ID&RA& Dec. & z & $L_{\rm IR}$ & HDFN ID & $f_{\rm
SX}$   & $f_{\rm HX}$   & $L_{\rm SX}$ & $L_{\rm HX}$ & HR & $f_{\rm HX}$/$f_{\rm R}$ &Spectral & X-ray\\
  & deg & deg &   & log\,(L$_\odot$) &         & erg/s/cm$^{2}$ &
erg/s/cm$^{2}$ & log (erg/s)     & log (erg/s)     &    &  & Class    & Class\\
(1)&(2)&(3)&(4)&(5)&(6)&(7)&(8)&(9)&(10)&(11)&(12)&(13)&(14)\\
 \hline
       1&189.28772&62.38866&1.26&12.39&---                &$<$2.37e-16&$<$7.67e-16&---&---&---&--- &---     &---\\
       2&189.27887&62.38520&0.50&11.42&---                &$<$1.39e-16&$<$2.98e-16&---&---&---&--- &---     &---\\
       3&189.29427&62.37627&1.52&12.39&---                &$<$1.34e-16&$<$3.39e-16&---&---&---&--- &---     &---\\
       4&189.27174&62.33079&0.84&11.75&---                &$<$7.65e-17&$<$8.15e-16&---&---&---&--- &---     &---\\
       5&189.35646&62.32817&0.28&10.75&hdfn-358            &1.86e-16&$<$3.93e-16&40.64&---&-1&---  &SFG      &SFG\\
       6&189.14424&62.32388&0.87&11.72&hdfn-086            &1.62e-16&9.04e-15&41.75&43.50&0.83$^{0.86}_{0.81}$&0.60$^{+0.03}_{-0.03}$&SFG      &AGN\\
       7&189.08646&62.31671&---&---&---                &$<$5.26e-17&$<$2.79e-16&---&---&---&---&---      &---\\
       8&189.54642&62.30510&---&---&---                &$<$2.44e-16&$<$7.17e-16&---&---&---&---   &---   &---\\
       9&189.22009&62.30213&0.25&10.90&---                &2.05e-17&$<$1.94e-16&39.59&---&-1&---   &---   &SFG\\
      10&189.44516&62.29540&1.15&11.81&---                &$<$1.11e-16&$<$3.15e-16&---&---&---&---   &---   &---\\
      11&189.39382&62.28980&0.64&11.62&---                &4.24e-17&$<$1.24e-16&40.85&---&-1&---  &---    &SFG\\
      12&189.51512&62.28651&0.28&11.01&---                &$<$1.55e-16&$<$3.56e-16&---&---&---&---   &---   &---\\
      13&189.08741&62.28600&1.99&12.63&---                &$<$7.07e-17&$<$3.02e-16&---&---&---&---  &---    &---\\
      14&189.57885&62.28439&---&---&---                &$<$7.20e-17&$<$6.14e-16&---&---&---&---  &---    &---\\
      15&189.13544&62.28315&0.44&11.26&---               &$<$2.61e-17&$<$2.75e-16&---&---&---&---   &---   &---\\
      16&189.31910&62.27872&0.56&11.28&hdfn-264            &4.54e-17&1.95e-16&40.73&41.37&0.03$^{+0.28}_{-0.20}$&-1.18$^{+0.22}_{-0.78}$&HEL AGN  &AGN\\
      17&189.37984&62.27229&0.98&11.89&---               &$<$7.81e-17&$<$1.31e-16&---&---&---&---   &---   &---\\
      18&188.99886&62.26384&0.38&11.38&hdfn-335            &1.55e-16&$<$4.42e-16&40.86&---&-1&--- &SFG      &SFG\\
      19&189.07638&62.26402&2.00&12.70&hdfn-291            &3.01e-17&$<$3.87e-16&41.89&---&-1&--- &---     &SFG\\
      20&188.99144&62.26019&0.38&11.60&hdfn-503            &4.66e-17&$<$4.47e-16&40.34&---&-1&--- &---     &SFG\\
      21&189.06708&62.25379&2.58&12.91&hdfn-117            &2.01e-16&1.12e-15&42.98&43.72&0.24$^{+0.08}_{-0.07}$&0.69$^{+0.09}_{-0.10}$&---     &AGN\\
      22&189.48220&62.25186&0.19&10.53&hdfn-356            &4.61e-16&5.48e-18&40.66&38.74&-0.59$^{+0.13}_{-0.14}$&-3.85$^{+1.26}_{-0.56}$&---     &SFG\\
      23&188.96211&62.24923&0.39&11.03&---                &$<$2.65e-16&$<$1.41e-15&---&---&---&---   &---   &---\\
      24&189.14271&62.24249&0.52&11.20&---                &$<$2.49e-17&$<$1.77e-16&---&---&---&---    &---  &---\\
      25&189.14832&62.23999&2.01&12.88&hdfn-082            &1.56e-16&1.73e-15&42.61&43.65&0.37$^{+0.07}_{-0.07}$&0.25$^{+0.05}_{-0.05}$&HEL AGN  &AGN\\
      26&189.53851&62.23786&---&---&---                &$<$1.08e-16&$<$7.02e-16&---&---&---&---     &---&--- \\
      27&189.26141&62.23376&1.25&12.03&---               &$<$3.21e-17&$<$1.14e-16&---&---&---&---     &---&--- \\
      28&189.15212&62.22800&0.56&11.09&---                &$<$2.06e-17&$<$1.08e-16&---&---&---&---     &---&--- \\
      29&188.97445&62.22703&0.88&12.00&---                &$<$6.34e-17&$<$3.07e-16&---&---&---&---     &---&--- \\
      30&189.50379&62.22669&0.44&11.77&hdfn-003            &3.72e-14&4.19e-14&43.41&43.46&-0.60$^{+0.01}_{-0.01}$&-0.31$^{+0.01}_{-0.01}$ &BL AGN&AGN\\
      31&189.29727&62.22530&2.00&12.82&hdfn-173            &5.75e-17&3.56e-16&42.17&42.96&0.15$^{+0.15}_{-0.14}$&-0.45$^{+0.11}_{-0.13}$ &---     &AGN\\
      32&189.37812&62.21623&---&---&---                &$<$4.86e-17&$<$1.43e-16&---&---&---&---     &---&--- \\
      33&189.08131&62.21457&0.47&11.52&hdfn-112            &8.15e-17&$<$5.19e-16&40.82&---&-1 &--- &HEL AGN  &AGN\\
      34&189.42242&62.21422&1.60&12.52&hdfn-408            &5.57e-17&5.03e-16&41.93&42.88&0.37$^{+0.19}_{-0.18}$&-0.30$^{+0.15}_{-0.21}$&SFG      &AGN\\
      35&189.14381&62.21139&1.22&12.56&hdfn-087            &6.15e-17&1.87e-16&41.69&42.17&-0.23$^{+0.15}_{-0.19}$&-0.86$^{+0.15}_{-0.18}$&SFG      &Amb\\
      36&189.14571&62.20672&0.56&11.30&---               &$<$3.01e-17&$<$9.47e-17&---&---&---&---     &---&--- \\
      37&189.42151&62.20581&---&---&---                &$<$5.30e-17&$<$1.83e-16&---&---&---&---     &---&--- \\
      38&189.43598&62.20522&0.23&10.41&---                &$<$1.56e-16&$<$1.65e-16&---&---&---&---     &---&--- \\
      39&189.14363&62.20360&0.46&11.65&hdfn-088            &1.47e-16&2.53e-16&41.04&41.28&-0.38$^{+0.10}_{-0.14}$&-2.10$^{+0.12}_{-0.14}$&HEL AGN  &AGN\\
      40&189.40440&62.20136&0.41&11.10&hdfn-493            &8.42e-17&$<$3.17e-16&40.69&---&-1&---    &---  &SFG\\
      41&189.42067&62.20013&1.17&12.06&hdfn-008            &7.26e-15&1.40e-14&43.71&44.00&-0.39$^{+0.02}_{-0.02}$&0.71$^{+0.02}_{-0.02}$&SFG     &AGN\\
      42&189.15318&62.19889&0.56&11.11&hdfn-080            &1.28e-15&2.20e-15&42.18&42.42&-0.48$^{+0.04}_{-0.03}$&-0.20$^{+0.04}_{-0.04}$&SFG      &AGN\\
      43&189.27452&62.19824&0.90&11.75&hdfn-413            &$<$4.08e-17&1.18e-17&---&40.65&1&-2.62$^{+0.89}_{-0.83}$&SFG      &Amb\\
      44&189.45862&62.19823&---&---&---                &$<$4.75e-17&$<$2.89e-16&---&---&---&---     &--- &---\\
     45&189.25661&62.19619&1.76&12.73&hdfn-471            &3.22e-17&$<$1.09e-16&41.79&---&-1&--- &---     &SFG\\
\hline
\hline
\end{tabular}
\end{table}
\addtocounter{table}{-1}
\begin{table}
\centering
\caption{- continued}
\begin{tabular}{|@{\hspace{-0.1mm}}l||@{\hspace{-0.1mm}}l||@{\hspace{-0.1mm}}l||@{\hspace{-0.1mm}}l||@{\hspace{-0.1mm}}l||@{\hspace{-0.1mm}}l||@{\hspace{-0.1mm}}l||@{\hspace{-0.1mm}}l||@{\hspace{-0.1mm}}l||@{\hspace{-0.1mm}}l||@{\hspace{-0.1mm}}l||@{\hspace{-0.1mm}}l||@{\hspace{-0.1mm}}l||@{\hspace{-0.1mm}}l|}
\hline
\hline
ID&RA& Dec. & z & $L_{\rm IR}$ & HDFN ID & $f_{\rm
SX}$   & $f_{\rm HX}$   & $L_{\rm SX}$ & $L_{\rm HX}$ & HR & $f_{\rm HX}$/$f_{\rm R}$ &Spectral & X-ray\\
  & deg & deg &   & log\,(L$_\odot$) &         & erg/s/cm$^{2}$ &
erg/s/cm$^{2}$ & log (erg/s)     & log (erg/s)     &    &  & Class    & Class\\
(1)&(2)&(3)&(4)&(5)&(6)&(7)&(8)&(9)&(10)&(11)&(12)&(13)&(14)\\

\hline
      46&189.03671&62.19546&1.34&12.43&---                &3.86e-17&$<$3.22e-16&41.58&---&-1&---  &---    &SFG\\
      47&189.05208&62.19458&0.28&10.79&---                &$<$5.80e-16&$<$3.53e-15&---&---&---&---   &---   &---\\
      48&189.22239&62.19434&1.27&12.30&hdfn-056            &9.08e-17&$<$1.66e-16&41.90&---&-1&--- &SFG      &SFG\\
      49&189.32612&62.19255&---&---&---                &$<$4.22e-17&$<$3.02e-16&---&---&---&--- &---     &---\\
      50&188.88702&62.18744&---&---&---                &$<$7.02e-16&$<$8.41e-16&---&---&---&---&---      &---\\
    51&189.01356&62.18636&0.64&11.74&hdfn-299            &1.42e-16&$<$5.94e-16&41.37&---&-0.54$^{+0.24}_{-0.22}$&--- &SFG      &SFG\\
      52&188.94873&62.18273&0.45&11.35&---                &$<$2.04e-16&$<$6.08e-16&---&---&---&--- &---     &---\\
      53&189.40081&62.18282&---&---&---                &$<$2.49e-16&$<$1.05e-15&---&---&---&--- &---     &---\\
      54&189.39812&62.18229&---&---&hdfn-431            &6.44e-17&5.89e-16&---&---&0.25$^{+0.24}_{-0.17}$&-0.57$^{-0.44}_{-0.75}$&---     &AGN\\
      55&189.00195&62.18134&---&---&---                &$<$9.20e-17&$<$2.07e-16&---&---&---&--- &---     &---\\
      56&189.12131&62.17943&1.01&12.07&hdfn-098            &1.62e-16&1.60e-15&41.91&42.91&0.37$^{+0.07}_{-0.08}$&0.32$^{+0.06}_{-0.06}$&---     &AGN\\
      57&188.97516&62.17871&0.51&11.19&---                &$<$6.25e-16&$<$2.90e-15&---&---&---&---  &---    &---\\
      58&189.36538&62.17665&0.21&10.50&---                &1.21e-16&$<$3.40e-16&40.20&---&-1&--- &---     &SFG\\
      59&189.21302&62.17524&0.41&11.25&hdfn-196            &8.44e-17&$<$2.12e-16&40.69&---&-1&--- &SFG      &SFG\\
      60&189.06332&62.16909&1.03&11.82&hdfn-438            &2.65e-17&5.91e-17&41.14&41.49&-0.01$^{0.35}_{-0.27}$&-1.47$^{+0.61}_{-1.41}$&---     &Amb\\
      61&189.14032&62.16832&1.02&11.93&hdfn-091            &2.66e-16&3.84e-16&42.13&42.29&-0.46$^{+0.10}_{-0.10}$&-0.78$^{+0.11}_{-0.12}$ &SFG      &AGN\\
      62&189.35540&62.16835&0.41&11.06&---                &$<$4.61e-17&$<$1.62e-16&---&---&---&---  &---    &---\\
      63&189.13030&62.16607&---&---&hdfn-212            &8.25e-17&5.74e-18&---&---&-0.46$^{+0.25}_{-0.22}$&--- &---    &---\\
      64&189.02738&62.16433&0.64&11.66&hdfn-405            &6.54e-18&1.61e-15&40.03&42.42&0.72$^{+0.11}_{-0.09}$&-0.27$^{+0.06}_{-0.07}$&SFG      &AGN\\
      65&189.23243&62.15487&0.42&11.17&---                &3.59e-17&$<$2.83e-16&40.34&---&-1&--- &---     &SFG\\
      66&189.34521&62.15184&---&---&---                &$<$6.40e-17&$<$1.08e-16&---&---&---&---   &---  &---\\
      67&189.29063&62.14482&0.90&12.27&hdfn-480            &2.11e-16&$<$7.28e-16&41.90&---&-1&--- &SFG      &SFG\\
      68&189.10174&62.14340&0.95&11.86&---               &6.05e-17&$<$2.81e-16&41.42&---&-1&---   &---   &SFG\\
      69&189.13844&62.14298&0.93&11.90&hdfn-093            &2.57e-15&3.21e-15&43.03&43.12&-0.56$^{+0.03}_{-0.03}$&-0.18$^{+0.04}_{-0.04}$&HEL AGN  &AGN\\
      70&189.19445&62.14258&0.97&12.11&hdfn-484            &5.52e-17&$<$1.77e-16&41.40&---&-1&--- &SFG      &SFG\\
      71&189.23308&62.13562&0.79&11.97&hdfn-328            &8.00e-17&$<$1.16e-16&41.35&---&-1&--- &SFG      &SFG\\
      72&189.22781&62.13450&0.84&11.65&---                &$<$2.38e-17&$<$1.06e-16&---&---&---&---     &---&--- \\
      73&189.26708&62.13202&1.25&12.24&hdfn-043            &2.73e-15&4.14e-15&43.36&43.54&-0.49$^{+0.03}_{-0.03}$&-0.07$^{+0.03}_{-0.03}$&HEL AGN  &AGN\\
      74&189.19095&62.13173&1.43&12.27&hdfn-275            &1.19e-16&$<$4.16e-16&42.14&---&-1&--- &SFG      &SFG\\
      75&189.27819&62.12293&---&---&hdfn-181            &6.86e-16&2.78e-15&---&---&-0.05$^{0.01}_{-0.11}$&0.45$^{+0.06}_{-0.06}$&---    &AGN\\
      76&189.08724&62.12061&1.15&12.14&---                &$<$9.39e-17&$<$6.36e-16&---&---&---&---     &---&--- \\
      77&189.14513&62.11948&0.63&11.35&---                &$<$6.57e-17&$<$1.43e-15&---&---&---&---     &---&--- \\
      78&189.15154&62.11859&0.28&10.67&---                &5.44e-17&2.49e-15&40.10&41.76&0.80$^{+0.85}_{-0.72}$&-0.97$^{+0.04}_{-0.04}$&---     &AGN\\
      79&189.06666&62.11690&---&---&---                &$<$1.28e-16&$<$6.32e-16&---&---&---&---    &---  &---\\
      80&189.13248&62.11211&---&---&---                &5.66e-18&$<$5.69e-16&---&---&-1&---&---      &---\\
      81&189.21462&62.11217&0.84&11.72&---                &8.67e-17&$<$3.49e-16&41.45&---&-1&--- &---     &SFG\\
      82&189.12056&62.10446&1.26&12.18&hdfn-213            &7.56e-16&7.44e-15&42.81&43.81&0.36$^{+0.07}_{-0.06}$&0.34$^{+0.04}_{-0.04}$&SFG      &AGN\\
      83&189.22622&62.09676&0.56&11.53&---                &$<$4.31e-17&$<$2.16e-16&---&---&---&---&---      &---\\
      84&189.14024&62.07928&0.52&11.46&---                &$<$7.00e-16&$<$2.29e-15&---&---&---&--- &---     &---\\
 \hline
 \hline
\end{tabular}
\label{table:all}
\end{table}
\twocolumn

\section*{Acknowledgments}
MS is grateful for MSSL/UCL support. The team is also extremely grateful to H\'{e}ctor Ot{\'{\i}}-Floranes, Kazushi Iwasawa and Bret Lehmer, for useful discussions and for providing us with models and data essential for the various comparisons made in this paper. 
This paper uses data from \textit{Herschel}'s submm photometer SPIRE. SPIRE has been developed by a consortium of institutes led by Cardiff
Univ. (UK) and including Univ. Lethbridge (Canada); NAOC (China); CEA, LAM
(France); IFSI, Univ. Padua (Italy); IAC (Spain); Stockholm Observatory
(Sweden); Imperial College London, RAL, UCL-MSSL, UKATC, Univ. Sussex
(UK); Caltech, JPL, NHSC, Univ. Colorado (USA). This development has been
supported by national funding agencies: CSA (Canada); NAOC (China); CEA,
CNES, CNRS (France); ASI (Italy); MCINN (Spain); SNSB (Sweden); STFC (UK);
and NASA (USA).

\bibliographystyle{mn2e}
\bibliography{references}

\label{lastpage}

\end{document}